\documentclass{aa}
\usepackage{natbib}
\bibpunct{(}{)}{;}{a}{}{,} 
\usepackage{graphicx}
\usepackage{txfonts}
\usepackage[normalem]{ulem}

\begin{document}

   \title{The life cycle of radio galaxies \\ in the LOFAR Lockman Hole field}

   \author{N. Jurlin\inst{1,}\inst{2}\thanks{jurlin@astro.rug.nl},
          R. Morganti\inst{1,}\inst{2},
          M. Brienza\inst{3,}\inst{4},
          S. Mandal\inst{5},
          N. Maddox\inst{6},
          K. J. Duncan\inst{5,}\inst{7},
          S. S. Shabala\inst{8,}\inst{9},
          M. J. Hardcastle\inst{9},
          I. Prandoni\inst{4},
          H. J. A. R\"ottgering\inst{5},
          V. Mahatma\inst{9},
          P. N. Best\inst{7},
          B. Mingo\inst{10},
          J. Sabater\inst{7},
          T. W. Shimwell\inst{2,}\inst{5},
          C. Tasse\inst{11,}\inst{12}
          }
\titlerunning{The life cycle of radio galaxies in the LOFAR Lockman Hole field}
   \institute{Kapteyn Astronomical Institute, University of Groningen, PO Box 800, 9700 AV, Groningen, The Netherlands
   \and ASTRON, Netherlands Institute for Radio Astronomy, Oude Hoogeveensedijk 4, 7991 PD, Dwingeloo, The Netherlands
   \and Dipartimento di Fisica e Astronomia, Università di Bologna, Via P. Gobetti 93/2, I-40129, Bologna, Italy
   \and INAF - Istituto di Radio Astronomia, Via P. Gobetti 101, I-40129 Bologna, Italy
        \and Leiden Observatory, Leiden University, P.O. Box 9513, 2300 RA Leiden, The Netherlands
   \and Faculty of Physics, Ludwig-Maximilians-Universit\"at, Scheinerstr. 1, 81679 Munich, Germany
  \and SUPA, Institute for Astronomy, Royal Observatory, Blackford Hill, Edinburgh, EH9 3HJ, UK
  \and School of Natural Sciences, Private Bag 37, University of Tasmania, Hobart, TAS 7001, Australia
  \and Centre for Astrophysics Research, School of Physics, Astronomy and Mathematics, University of Hertfordshire, College Lane, Hatfield AL10 9AB, UK
  \and School of Physical Sciences, The Open University, Walton Hall, Milton Keynes, MK7 6AA, UK
  \and GEPI, Observatoire de Paris, Université PSL, CNRS, 5 Place Jules Janssen, 92190 Meudon, France
  \and Department of Physics \& Electronics, Rhodes University, PO Box 94, Grahamstown, 6140, South Africa}

    \authorrunning{Jurlin et al.}
  \date{\today}
   \abstract
   {Radio galaxies are known to go through cycles of activity, where phases of apparent quiescence can be followed by repeated activity of the central supermassive black hole. A better understanding of this cycle is crucial for ascertaining the energetic impact that the jets have on the host galaxy, but little is known about it.
   We used deep LOFAR images at 150 MHz of the Lockman Hole extragalactic field to select a sample of 158 radio sources with sizes $> 60^{\prime\prime}$ in different phases of their jet life cycle. Using a variety of criteria (e.g. core prominence combined with low-surface brightness of the extended emission and steep spectrum of the central region) we selected a subsample of candidate restarted radio galaxies representing between 13\%\ and 15\%\ of the 158 sources of the main sample. We compare their properties to the rest of the sample, which consists of remnant candidates and active radio galaxies. Optical identifications and characterisations of the host galaxies indicate similar properties for candidate restarted, remnant, and active radio galaxies, suggesting that they all come from the same parent population.
   The fraction of restarted radio galaxies is slightly higher with respect to remnants, suggesting that the restarted phase can often follow after a relatively short remnant phase (the duration of the remnant phase being a few times 10$^{7}$ years). This confirms that the remnant and restarted phases are integral parts of the life cycle of massive elliptical galaxies. A preliminary investigation does not suggest a strong dependence of this cycle on the environment surrounding any given galaxy.}

\authorrunning{The life cycle of radio galaxies in the LOFAR Lockman Hole field}
\titlerunning{Jurlin et al.}
\maketitle

\section{Introduction}\label{sec:introduction}
The energy released by an active black hole can have a significant effect on the evolution of its host galaxy. Winds, radiation, and radio-emitting plasma jets can produce outflows that expel gas from the bulge, or that can prevent hot gas from cooling and forming stars. These processes are considered to potentially be able to affect the evolution of the host galaxy (\citealt{2018NatAs...2..198Harrison} and references therein).
However, they can only be fully effective if they are recurrent in the life of the host galaxy (\citealt{2009ApJ...699..525ShabalaAlexander,2010ApJ...717..708Ciotti,2015A&A...579A..62Gaspari,2017NatAs...1..596Morganti,2017MNRAS.471..658Raouf}).
Radio-loud active galactic nuclei (AGN) emit energy via jets and are known to go through phases of activity. Therefore, radio AGN activity can provide a mechanism to regulate accretion or cooling of the surrounding gas and possibly the star formation rate (SFR) and the growth of the supermassive black hole (SMBH, \citealt{2014ARA&A..52..589HeckmanBest_review}).
It is therefore essential to quantify the duty-cycle of this activity to understand its impact.
The current understanding of how long this cycle lasts and what physical properties it depends on is still incomplete.

For radio-loud AGN, different phases of their life-cycle can be identified. These phases include young, {\sl newly born} radio sources (which are thought to be represented by gigahertz peak spectrum (GPS) and compact steep spectrum (CSS) sources; \citealt{1998PASP..110..493Odea, 2012ApJ...760...77An_Baan, 2016AN....337....9Orienti}), and evolved sources in an active phase, which can be followed by a {\sl remnant phase} when the activity stops or substantially decreases (e.g. \citealt{1994A&A...285...27KomissarovGubanov,2007A&A...470..875Parma2007,2011A&A...526A.148Murgia2011,2015A&A...583A..89Shulevski2015,2017A&A...606A..98B,2018MNRAS.475.4557Mahatma_remnants}). After this remnant phase, a {\sl restarted phase} can occur, where the activity reignites after the central engine has gone through a period of quiescence or low activity. This latter phase is particularly relevant for feedback, and indeed the present study. 

Statistical studies using luminosity functions (\citealt{2005MNRAS.362...25Best_luminosity, 2008MNRAS.388..625Shabala, 2019A&A...622A..17Sabater}) provide constraints on the life-cycle of radio-loud AGN and show its dependence on the radio luminosity and on the stellar mass of the host galaxy.
According to these studies, radio galaxies become active in a high-power phase at intervals of between one and a few gigayears, while they would need to spend more than a quarter of their life active in a low-power phase (\citealt{2005MNRAS.362...25Best_luminosity}).
\citet{2019A&A...622A..17Sabater} concluded that the most massive galaxies (>$\rm 10^{11}$ $M_{\sun}$) are always switched on at some level, in particular at low radio luminosities (log ($L_{\rm 150~MHz}/\rm W~Hz^{-1}$) $\geq 21.7$).

Short outburst intervals in the range 1-10 Myr (see \citealt{2014MNRAS.442.3192Vantyghem} and references therein) are also derived from studying the structures and size of multiple X-ray cavities in the intracluster medium of nearby clusters. One scenario says that these cavities are generated by multiple phases of jet activity (sometimes already invisible at radio wavelengths) affecting the intergalactic medium (see \citealt{2012NJPh...14e5023Mcnamara12} and references therein).

To confirm these results we need to identify candidate restarted radio sources using the morphology and spectral properties of the radio emission. The fraction of these sources holds the key to understanding the length of the jet duty cycle.

Recognising a restarted radio source is not straightforward as it may have a variety of properties depending on the evolution of the radio source itself.
One well-known population of restarted objects is the `double-double' radio galaxies (DDRG; \citealt{2000MNRAS.315..371Schoenmakers}). These sources contain two pairs of distinct lobes on opposite sides of the host galaxy.
The fact that we can see multiple pairs of lobes implies that the time required for the jets to restart is shorter than that for the outer lobes to fade past detection in radio images. Based on models of spectral ageing, the duration of the quiescent phase of DDRGs is typically in the range of $10^5$ - $10^7$ yr and is never more than 50\%\ of the duration of the previous active phase, which is of the order of $\sim$ $10^8$ yr (\citealt{2013MNRAS.430.2137Konar}).
Although rare, more than two episodes of jet forming activity have been observed (\citealt{2007MNRAS.382.1019Brocksopp, 2011MNRAS.417L..36Hota}).

However, DDRGs represent only one of the possible ways in which the restarted activity can manifest itself.
In some cases, a new episode of activity is indicated by the presence of a GPS/CSS source inside the nuclear region, implying the presence of compact, newly formed jets (e.g. \citealt{1974Natur.250..625Willis, 1985A&A...148..243Barthel, 1998PASP..110..493Odea, 2005A&A...443..891Stanghellini, 2010ApJ...715..172Tremblay, 2012A&A...545A..91Shulevski}; \citealt{2019ApJ...875...88Bruni}).

In at least one case (3C388), restarted activity has been identified using the information on the spectral index distribution within the radio lobes (\citealt{1982ApJ...257..538Burns, 1994ApJ...421L..23Roettiger, 2020arXiv200313476Brienza388}). Based on single-frequency morphology alone, this source would be classified as a `normal' active source. 

So far, the only attempt to perform a blind selection was presented by \citet{2012ApJS..199...27Saripalli}. In particular, these latter authors searched for the presence of low-surface brightness (SB, with SB < 10 mJy arcmin$^{-2}$ at 1400 MHz) structures as an indication of a previous cycle of activity. 
They present classification criteria based on the morphology of the radio emission, separating Fanaroff-Riley type I (FRI; \citealt{1974MNRAS.167P..31Fanaroff}) and type II (FRII) radio galaxies. 
\citet{2012ApJS..199...27Saripalli} derive a fraction of remnant sources (i.e. with no signature of ongoing nuclear activity) of 3\%\ and a larger fraction of 24\%\ (33\%\ FRII and 13\%\ FRI) showing signatures of restarted radio emission. They conclude that the remnant phase in both types of radio galaxies is relatively short. FRII sources may spend approximately two-thirds of their lifetime in the active phase, one-third in the restarting phase, and only a relatively short time in the dying phase, while in the case of FRI sources the active phase may be longer and the restarting phase correspondingly shorter. 

Another important aspect to be considered is the possible relation between restarted activity and the properties of the host galaxy. This may provide insight into whether or not there is a relation between the host galaxy properties and the jets phase.
A few studies, focused only on DDRG, have covered this aspect with some discordant results (see e.g. \citealt{2012BASI...40..121NandiSaikia}, \citealt{2017MNRAS.471.3806Kuzmicz} and \citealt{2019A&A...622A..13Mahatma_restarted}). \citet{2017MNRAS.471.3806Kuzmicz} found that the stellar masses of the host galaxies of DDRGs are lower compared to those of the evolved active population. However, their study is mostly based on a collection of objects from the literature, and therefore the sample is likely inhomogeneous. On the other hand, \citet{2019A&A...622A..13Mahatma_restarted}, using 33 morphologically selected DDRGs, found no significant difference between hosts of DDRGs and the parent population of radio-loud AGN indicating that they both belong to the same population of host galaxies.
As a consequence, these latter authors propose that the restarted activity might be caused by changes in the mass accretion rate on short timescales or by variation of the accreted magnetic flux density. 
 
Based on the above, there is a clear requirement to create a larger and more representative sample of candidate restarted radio galaxies. This is now possible thanks to the new low-frequency radio surveys, in combination with ancillary data, providing the possibility to expand the selection and identification of candidate restarted radio sources using a broader range of criteria. 
The main goal of this paper is to carry out such a selection using a combination of quantitative and comprehensive criteria (described in Sect.~\ref{sec:scenarios_and_selection}) and compare the fraction of these objects with the fraction of sources in other phases, namely remnant and active.

The approach we take here is to focus on one of the well-studied extragalactic fields, the Lockman Hole (LH, \citealt{1986ghg..conf...75Jahoda}) where we create a sample of sources with angular sizes $> 60^{\prime\prime}$ among which we identify candidate restarted radio galaxies. We complement this with the selection of candidate remnant radio sources in the same field presented by \citet{2017A&A...606A..98B}. All the other sources in the sample, which are not classified as remnant or restarted candidates, are considered as presently active sources, and we use these as a comparison sample in our analysis.
This provides the possibility to compare the fraction of restarted, remnant, and active radio galaxies, key information for timing their life cycles. Furthermore, optical identifications were made for all the radio sources in the sample, allowing us not only to compare the radio properties of the sources in these three groups but also to compare the properties of the host galaxies. In a companion paper (Shabala et al., accepted) we use our results as input for theoretical models and discuss the implications for the radio galaxy life-cycle.

The paper is structured as follows. In Sect.~\ref{sec:scenarios_and_selection} we describe the criteria we use to select the candidate restarted radio galaxies. 
The description of the LH data, which includes the radio data and ancillary optical and infrared data, is given in Sect.~\ref{sec:LH_data}. The construction of the initial sample used for the application of the criteria and the optical identification are described in Sect.~\ref{sec:sample_selection}. In the same section, we present details of the criteria we use to select candidate restarted radio galaxies and results from the selection. Section~\ref{sec:derived_properties_of_the_host_galaxy} introduces the host galaxy properties, stellar masses, WISE colours, and the environment. 
We also studied the radio properties (linear sizes and radio luminosities) of the sources in the sample and we discuss those in Sect.~\ref{subsec:discussion_radio_properties}.
In Sect.~\ref{sec:discussion}, we discuss the results of the selection of restarted candidates, derived properties of the host galaxy, and the radio properties for our complete sample. 
A summary and conclusions follow in Sect.~\ref{sec:summary_and_conclusions}.

The cosmology adopted throughout the paper assumes a flat universe and the following parameters: $\rm H_{0} = 70$ $\rm km$ $\rm s^{-1}$ $\rm Mpc^{-1}$, $\rm \Omega_{\Lambda} =0.7$, $\rm \Omega_{M} =0.3$. The spectral index $\alpha$ is defined as $S_{\nu} \propto \nu^{-\alpha}$.

\section{Criteria for selecting candidate restarted radio sources}\label{sec:scenarios_and_selection}

Here we describe the criteria used for the selection of candidate restarted radio galaxies and provide our motivation for choosing them. We also briefly present the considerations that led us to the adopted approach. This represents a key step in this study.
We tried to take an approach to selecting the
candidate restarted radio sources that is as automated as possible. 

\begin{figure}
\centering
\minipage{\textwidth}
  \includegraphics[width=0.5\linewidth]
  {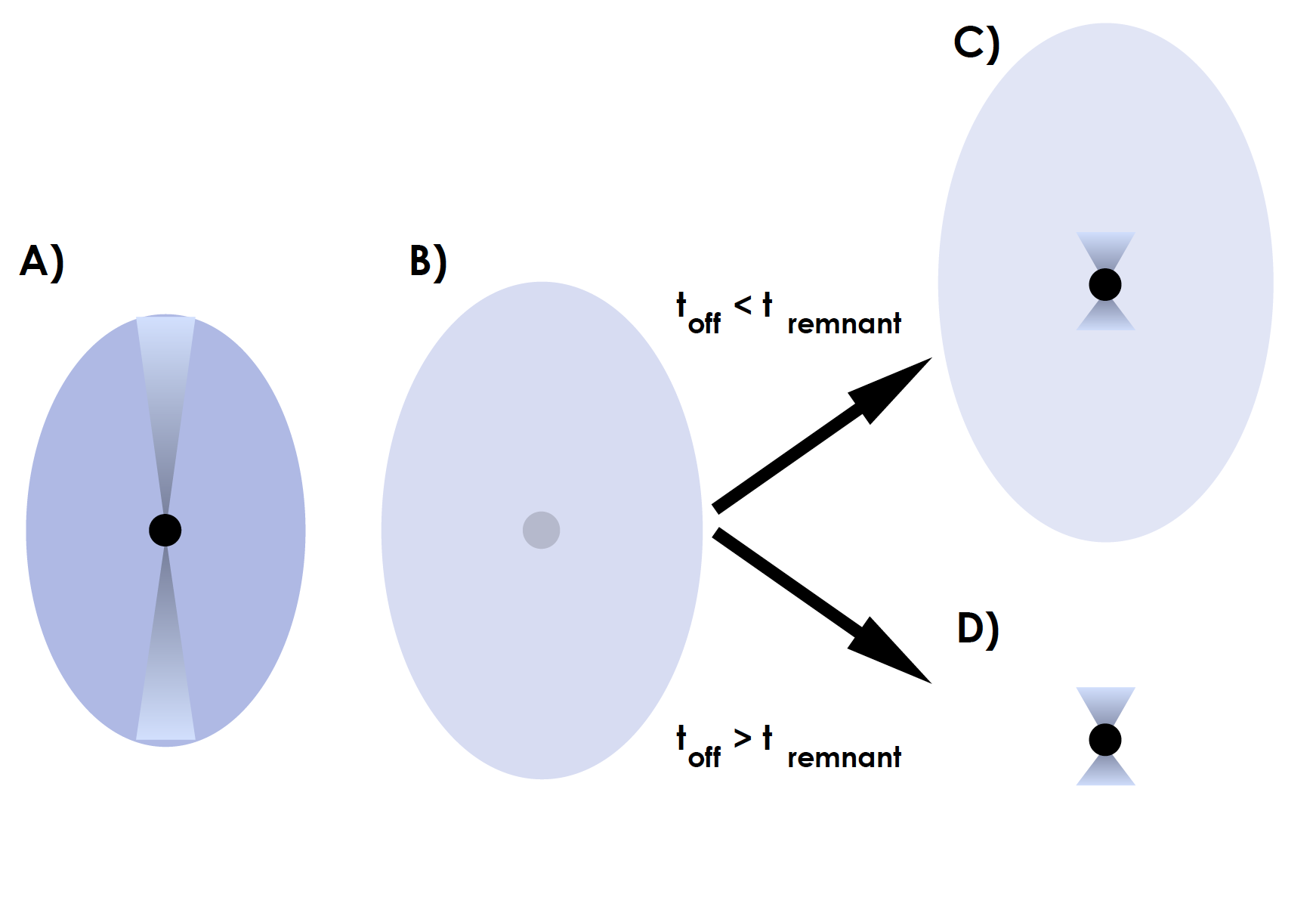}
\endminipage
\caption{Sketch of the life cycle of radio galaxies with different scenarios for the restarted phase. Here we show an active galaxy (A) going through a remnant phase (B), and then through a restarted phase (C and D). Depending on the duration of the active phase with respect to the remnant fading time we can observe a restarted morphology (C) or a simple compact (young) radio source (D).}
\label{fig:off_time_scenarios_sketch}
\end{figure}

The possible evolutionary scenarios that we assume are illustrated in Fig.~\ref{fig:off_time_scenarios_sketch}.
We consider a source where the nuclear activity switches off or dims and the lobes become remnant structures (Fig.~\ref{fig:off_time_scenarios_sketch}B).
In this phase, the source is characterised by low-SB and amorphous structures, and the cores are invisible or very weak (see e.g. \citealt{2017A&A...606A..98B} and \citealt{2018MNRAS.475.4557Mahatma_remnants}).
If a new phase of jet activity follows on a relatively short timescale ($t_{\rm off}<t_{\rm remnant}$, where $t_{\rm off}$ is the time in which the jets are not active and $t_{\rm remnant}$ is the time before the remnant lobes become too faint to be detected), we expect the morphology shown in Fig.~\ref{fig:off_time_scenarios_sketch}C. In this case, we expect to see a relatively bright radio "core" surrounded by a diffuse, low-SB, and amorphous structure.
If on the contrary, $t_{\rm off}>t_{\rm remnant}$ (Fig.~\ref{fig:off_time_scenarios_sketch}D), we expect to detect only a young radio source and will not be able to classify it as a restarted radio galaxy.
With this scenario in mind, we therefore aim to detect restarting radio sources as shown in Fig.~\ref{fig:off_time_scenarios_sketch}C, using a combination of criteria.

The first criterion is based on core prominence. We select sources with high core prominence (CP) - compared to active sources - combined with low-SB extended emission.
We define {CP} as the ratio between the flux density of the core and the total flux density of the source, $\rm CP= S_{\rm core}\rm \slash{S_{\rm total}}$, where the core is defined as the central compact feature coinciding with the optical identification (see below and Sect.~\ref{sec:optical_identification}). In restarted radio galaxies we expect to observe a high CP due to the combination of fading of the total extended emission and the start of a new jet in the central region. This is similar to what can be seen in a number of cases from the literature (e.g. \citealt{2007MNRAS.378..581Jamrozy, 2012A&A...545A..91Shulevski, 2016A&A...592A..94Frank, 2018A&A...618A..45Brienza, 2019arXiv191204812Sridhar}), and is consistent with the rapid fading of remnants predicted by models (\citealt{2002MNRAS.336..649KaiserCotter2002, 2018MNRAS.475.2768Hardcastle,2018MNRAS.474.3361TurnerRAISEIII} and references therein).
A similar criterion was also used by \citet{2012ApJS..199...27Saripalli}. However, unlike in this latter study, where high CP sources were selected by visual inspection, we automatically apply a CP cut to our sample; see Sect.~\ref{subsec:results_CP}.

The second criterion is based on spectral index. We select sources that have a steep spectrum ($\alpha^{\rm 150 MHz}_{\rm 1400 MHz}>0.7$, where $S_{\nu} \propto \nu^{-\alpha}$) inner region (`core').
This identifies sources where newly restarted {\sl bright} radio jets are present in the inner region but are unresolved by our observations. For the available resolution of 6$^{\prime\prime}$ (see Sect.~\ref{sec:LH_data}), this corresponds to a linear size between 6 and 40 kpc for a range in redshift between 0.05 and 0.7, typical of our sources; see Sect.~\ref{subsec:discussion_redshiftGroups}. These sizes are typical of CSS sources \citep{1998PASP..110..493Odea} which are expected to have $\alpha$ > 0.7 in our frequency range. This criterion has been used in other studies to select candidate restarted radio galaxies (e.g. 3C~293, \citealt{1981AJ.....86.1294Bridle} and 3C~236, \citealt{1981ESASP.162..107Schilizzi}).

After applying the aforementioned criteria, a final visual inspection of the restarted candidates is necessary to verify the reliability of their automatic classification. In this way, we can reject sources that appear to have collimated jet-like structures indicating ongoing fuelling of the large-scale lobes and/or where the high CP appears to be the result of beaming effects.

Furthermore, visual inspection of all the sources in the LH sample can allow us to identify sources like DDRGs and, in general, restarted sources which were not selected in the automatic way described
above but where the inner jets are resolved.

We note that, despite the use of a variety of selection criteria, selecting all classes of restarted radio galaxies remains challenging. For example, sources showing signs of restarting activity in their spatially resolved spectral properties (as in the case of 3C~388; \citealt{1982ApJ...257..538Burns, 1994ApJ...421L..23Roettiger,2020arXiv200313476Brienza388}) might be missed.

Finally, we point out that the appearance of restarted sources does not depend only on $t_{\rm off}$ and $t_{\rm remnant}$,
but there are other parameters that influence this cycle, such as the magnetic field (see Sect.~\ref{sec:discussion_occurence}), environment \citep{2013MNRAS.430..174HardcastleKrause13,2015ApJ...806...59TurnerShabala,2018MNRAS.480.5286Yates,2019MNRAS.482..240Krause}, and the properties of the plasma \citep{2018MNRAS.476.1614Croston}. Taking all these factors into account requires more detailed data than what are available at this stage.

\section{Lockman Hole data} \label{sec:LH_data}

Our study explores radio galaxies in the LH area.
The LH is one of the most thoroughly studied extragalactic regions of the sky centred at R.A. 10h 45m, Dec. $\rm +58^\circ$ $\rm 00^\prime$. Extensive multi-wavelength ancillary data are available in this field (see e.g. \citealt{2018MNRAS.481.4548Prandoni} for an overview). In particular, this is one of the fields for which there exist deep Low Frequency Array (LOFAR, \citealt{2013A&A...556A...2VanHaarlem}) observations covering $\sim$ 30 deg$^{\rm 2}$ (\citealt{2016MNRAS.463.2997Mahony}; Mandal et al., in prep and Tasse et al., in prep). The above observations make the LH an ideal region for investigation of classes of rare sources such as remnant or restarted radio galaxies. 

In order to apply the selection criteria described in the previous section, it is essential to use a variety of radio images with both good sensitivity to detect low-SB structures and high spatial resolution to identify compact components. In this way, we can obtain information on both the large-scale morphology and the presence of radio cores. 

For this work we used the published image and catalogue at 150~MHz presented by \citet{2016MNRAS.463.2997Mahony} with a resolution of 18$^{\prime\prime}$ and reaching a noise level of rms = 150 $\rm \mu$Jy beam$^{-1}$. 
In addition, we used a newly available and deeper high-resolution image presented by Mandal et al. (in prep) and Tasse et al. (in prep).
This image was obtained using an improved version of the calibration pipeline used to process the LOFAR Two-metre Sky Survey (LoTSS, see \citealt{2018A&A...611A..87Tasse,2019A&A...622A...1Shimwell}, Tasse et al., in prep), which takes into account direction-dependent effects and provides images with higher sensitivity and image fidelity, and a longer integration time of 96 hours. The image reaches a noise level of rms = 28 $\rm \mu$Jy beam$^{-1}$ with a spatial resolution of 6$^{\prime\prime}$.

In order to derive the radio properties of the sources, as well as the properties of their host galaxies, we carried out optical identifications of the final selected sample using the ancillary data available for the LH extragalactic field.

For the optical identifications we used the Sloan Digital Sky Survey Data Release 14 (SDSS DR14; \citealt{2018ApJS..235...42Abolfathi}) and the Wide-field Infrared Survey Explorer (WISE; \citealt{2010AJ....140.1868Wright,2011ApJ...731...53Mainzer}), which mapped the sky at 3.4, 4.6, 12, and 22 $\rm \mu$m (W1, W2, W3, W4) with an angular resolution of 6.1$^{\prime\prime}$, 6.4$^{\prime\prime}$, 6.5$^{\prime\prime}$, and 12.0$^{\prime\prime}$, in the four bands respectively. In particular, we used the AllWISE Source Catalogue (\citealt{2014yCat.2328....0Cutri14allWISE}), which is a significant improvement over the WISE All-Sky Release Catalogue (\citealt{2012yCat.2311....0CutriallSky}). AllWISE provides enhanced photometric sensitivity and accuracy, and improved astrometric precision compared to the WISE All-Sky Data Release. For the process of the optical identification see Sect.~\ref{sec:optical_identification}.

\section{Sample selection}\label{sec:sample_selection}

\subsection{Initial sample selection}
As mentioned in Sect.~\ref{sec:introduction}, the purpose of this work is to build a sample of radio galaxies in different phases of their evolution including remnant and restarted sources, while the remaining sources provide a comparison sample of active radio galaxies.

We limited our analysis to radio galaxies with angular sizes > 60$^{\prime\prime}$. This value was set following \citet{2017A&A...606A..98B} for the selection of remnant radio galaxies and motivated by the need to select sources covering at least three beams in the LOFAR 18$^{\prime\prime}$ map which allows for morphological classification. By applying the same cut, we can include this class of sources in our analysis and, thanks to the use of the LOFAR image at 6$^{\prime\prime}$ resolution, we can improve the morphological characterisation of these sources. For the median redshift of our sample, the angular size of 60$^{\prime\prime}$ corresponds to $\sim$ 350 kpc.
The selection was done from the low-resolution catalogue produced by \citet{2016MNRAS.463.2997Mahony}, which lists a total of 5323 sources or source components.
The catalogue was produced using Python Blob Detector and Source Finder (PyBDSF; \citealt{2015ascl.soft02007MohanRaffertyPYBDSF}), which does not provide a perfect representation of extended and complex radio sources. To overcome this, we visually inspected each detection with a PyBDSF angular size > 60$^{\prime\prime}$ and if necessary joined it with other components that were not automatically recognised as being part of the same source (see also \citealt{2019A&A...622A...2Williams} for a detailed discussion). In these cases, the total flux density of the source was recomputed by adding all components together, while the size was measured on the final image of the source inside 3$\sigma$ contours.
We then excluded five low-redshift spiral star-forming galaxies based on the optical SDSS counterpart, and ten sources strongly affected by calibration artefacts produced by nearby sources of high flux density.
The final sample used for our study consists of 158 visually confirmed radio galaxies. These numbers are also summarised in Table~\ref{tab:catalogue_description_process_table}.

\subsection{Optical identification} \label{sec:optical_identification}
\begin{figure} [h]
\includegraphics[width=\linewidth]{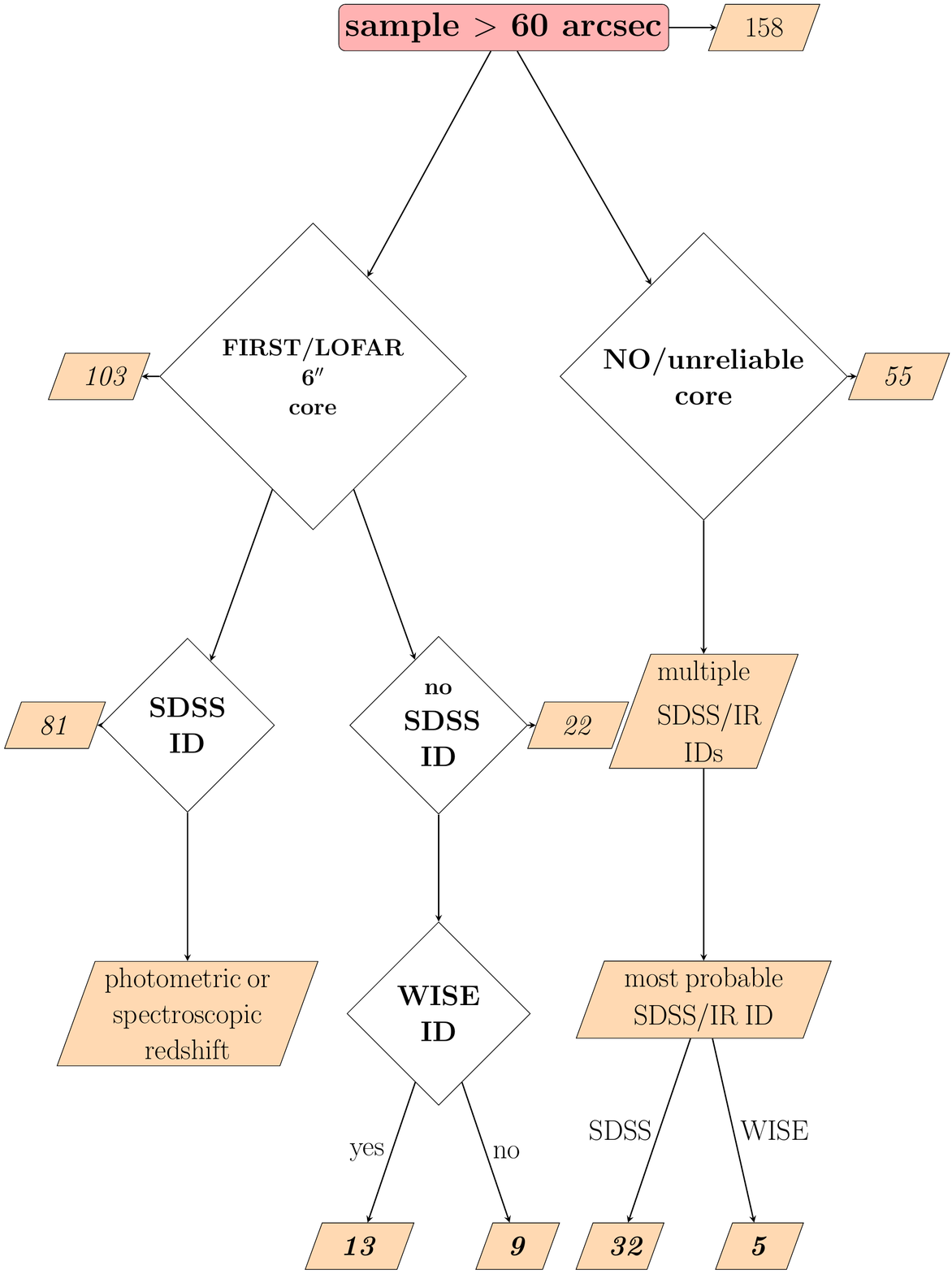}
\caption{Flow chart showing the process used to identify the host galaxy of the sources in the radio galaxy sample.}
\label{fig:flow_chart}
\end{figure}

The flow chart shown in Fig.~\ref{fig:flow_chart} summarises the strategy followed for the host galaxy identification process as described below.
For the sources where the radio core could be identified from the radio images, either from the high-resolution LOFAR 150 MHz 6$^{\prime\prime}$ image or the FIRST image at 1400 MHz, we obtained the optical identification using SDSS DR14. For the sources without a detection of the core, we considered all the optical counterparts based on the morphology of both the radio galaxy and the potential optical hosts. Morphology of the radio source meant that we decided on the most probable one based on the source's barycenter, while the morphology of the host galaxy would be a massive elliptical because those are the most probable radio host galaxies.
In the case of a reliable optical counterpart, we associated a redshift value to that source. We have redshift information for 113 of the 158 sources of our initial sample (i.e. 72\%\ of the sample). For 60 of these 113 sources (i.e. 53\% of the sources with optical IDs) the redshift is derived from SDSS spectra. For the remaining sources, the redshifts are photometric, also taken from SDSS DR14.
The distribution of redshifts is presented and discussed in Sect.~\ref{subsec:discussion_redshiftGroups}.

In case no counterpart was found in SDSS DR14 but was instead detected in the WISE image, we were able to put a constraint on the redshift of the source based on a modified version of the well-known K--z relation for radio galaxies \citep{2004A&A...415..931Rocca-Volmerange}. Using a flexible stellar population synthesis code (\citealt{2019ApJ...876..110DuncanCANDELS} and references therein) we calculated the apparent WISE W1 magnitude corresponding to an old, quiescent stellar population with a stellar mass of $\rm 10^{11}~M_{\sun}$ (consistent with the WISE W1 magnitudes observed for the sample with 
available redshifts). We were then able to infer an approximate redshift limit for a further 18 sources with or without the detection of the core (i.e. 11\%). However, we note that these redshift approximations assume that radio sources without redshifts are hosted in galaxies with the same underlying host galaxy properties. Removing these 18 sources from the analysis does not change the distributions and the results of statistical analyses.
The remaining sources, that is, those for which an estimate of redshift is not available, are not considered in the analysis of the optical and radio properties.
Table~\ref{tab:catalogue_description_process_table} summarises the number of objects in our sample, as well as the detection of the core in FIRST or LOFAR images and optical or infrared identifications.

\begin{table} [h]
\caption{Number of sources satisfying each criteria that led to the final catalogue used for the search for candidate restarted radio galaxies.}
\label{tab:catalogue_description_process_table} 
\resizebox{\linewidth}{!}{
\centering
\begin{tabular}{l l l}
\hline\hline
\textbf{Initial catalogue obtained by PyBDSF} & \citet{2016MNRAS.463.2997Mahony} \\ \hline
        N of detections & 5323  \\
        N of sources with angular size $>$ 60$^{\prime\prime}$ & 173  \\ \hline
\textbf{Catalogue with sources $>$ 60$^{\prime\prime}$} & \\ \hline
N of sources & 173 \\
N of sources not associated to AGN & 5 \\
N of sources affected by artefacts & 10 \\ \hline
 \textbf{Final catalogue ($>$ 60$^{\prime\prime}$)} & \\
\hline

 N of sources & 158 \\
N of sources with  core in FIRST & 68 
 & 43 \%\\
N of sources with  core in LOFAR 6$^{\prime\prime}$ & 86 
 &  54 \%\\
N of optical identifications & 113
 & 72 \% \\
N of WISE lower limits for the redshift & 18
 & 11 \% \\ \hline 
\end{tabular}
}
\end{table}

\subsection{Selection of restarted radio galaxies}
\label{sec:results_restarted_selection}

Here, we explore the selected sample of 158 sources to identify candidate restarted radio galaxies detailing the three criteria introduced in Sect.~\ref{sec:scenarios_and_selection} and the data presented in Sect.~\ref{sec:LH_data}.
Remnant candidates have already been selected from the same field and using the same cut in size by \citet{2017A&A...606A..98B}. This group of remnant candidates, which includes 18 sources present in our sample, is used in our analysis. We would like to point out that the number of remnants used for this work is smaller than the total number presented in \citet{2017A&A...606A..98B} (23 sources). The reason for this is that thanks to the new LOFAR image at 6$^{\prime\prime}$ resolution and a VLA follow up of the sources that will be described in a forthcoming paper (Jurlin et al. in prep) we were able to reject two sources as remnant candidates. Three more sources are located outside the LOFAR catalogue used to set up the original sample of sources in this work (either outside of the sky area covered or rejected because of their reported size).

\subsubsection{High radio core prominence and low surface brightness} \label{subsec:results_CP}

We first selected candidate restarted radio sources based on their high CP and low-SB extended emission.
To derive the CP$_{\rm 1400~MHz}$ of the sources in the sample, we used the total flux densities reported in the National Radio Astronomy Observatory (NRAO) Very Large Array (VLA) Sky Survey (NVSS; \citealt{1998AJ....115.1693Condon}). If the total emission was not detected in the NVSS catalogue, we inspected the images visually. If there was a detection below 5$\sigma_{\rm catalogue}$ (where $\sigma_{\rm catalogue}$ is the typical rms of 0.45 mJy reported in the NVSS catalogue) in the NVSS image, we measured the flux density directly from the image. For those with no detection in the NVSS image, we compute a 3$\sigma_{\rm local}$ upper limit by measuring the standard deviation of the flux density in ten different boxes surrounding the source location (see Table~\ref{tab:candidate_restarted_appendix}).
The core flux density, which is used to derive both spectral index and CP, was measured directly from the VLA Faint Images of the Radio Sky at Twenty-cm survey (FIRST; \citealt{1995ApJ...450..559BeckerFIRST}) images at 1400 MHz with an angular resolution of 5$^{\prime\prime}$.
Given the importance of a reliable estimate of the core flux density for our study, extra effort was made to visually inspect the FIRST images and to measure the values of the peak flux density of the core - if present - when above 3$\sigma_{\rm catalogue}$, where $\sigma_{\rm catalogue}$ is the typical rms of 0.15 mJy reported in FIRST catalogue. For the sources without a detected core in FIRST, we measured the local noise directly from the map ($\sigma_{\rm local}$, which can be lower than the noise value reported in the FIRST catalogue), and we assumed 3$\sigma_{\rm local}$ as the upper limit of the core flux density (see Table~\ref{tab:candidate_restarted_appendix}).
The CP$_{\rm 1400~MHz}$ was computed at 1400 MHz since at this frequency our derived values can be more easily compared to those obtained in other studies in the literature (e.g. \citealt{1990A&A...227..351DeRuiter, 2014MNRAS.437.3405LaingBridle} and \citealp{2015A&A...576A..38Baldi}).
Powerful radio galaxies (i.e. FRII) have relatively faint cores compared to their extended radio emission. FRI radio galaxies have higher CP values, while the most extreme CP values have been found in FR0 (see \citealt{2015A&A...576A..38Baldi}).
This trend is also illustrated by the anticorrelation between CP and radio luminosity found by \citet{1990A&A...227..351DeRuiter}.

The aforementioned studies show that CP $ > 0.1$ represents much higher values of CP than what is typically found in FRI radio sources, as the bulk of the radio galaxies in the well-known 3C and B2 catalogue show a CP < 0.1.
Therefore, we decided to assume this value as a selection criterion for the CP of our restarted candidates. 
However, we note (as shown in Table~\ref{tab:candidate_restarted}) that the majority of CP-selected candidates have a CP > 0.2, which is well above our cutoff limit.

Among all the sources with high CP, we only selected as restarted candidates those where the extended emission at 150 MHz had very low values of SB. We used the LOFAR 18$^{\prime\prime}$ image to identify and measure the low-SB emission. A value of SB$_{\rm 150~MHz}$ $\leq$ $\rm 50$ $\rm mJy$ $\rm arcmin^{-2}$ was used, following \citet{2017A&A...606A..98B}. This value is also consistent with what has been used by \citet{2012ApJS..199...27Saripalli}, namely SB $< 10$ mJy arcmin$^{-2}$ at 1400 MHz. At the median redshift of our sample ($ z \sim 0.4$) this corresponds to a minimum luminosity of 6 $\rm \times$ 10$\rm ^{24}$ W Hz$^{-1}$ at 1400 MHz. Models (\citealt{2017MNRAS.471..891Godfrey,2018MNRAS.475.2768Hardcastle, 2018MNRAS.474.3361TurnerRAISEIII} and references therein; Shabala et al., accepted) show that remnant sources fade rapidly, and hence brighter lobe emission is likely to correspond to active sources. As described in Sect.~\ref{sec:scenarios_and_selection}, in this way we are more likely to select extended emission that represents a remnant emission from a previous phase of jet activity.

As shown in Table~\ref{tab:catalogue_description_process_table}, for 57\% of the sources in our sample we only have an upper limit to the core flux density at 1400 MHz with FIRST (43\% FIRST detections). 
For the sources in this group that also have a total flux density below 10 mJy at 1400 MHz, the derived CP upper limit (higher than 0.1) does not provide a useful constraint for the CP. Therefore, we exclude these sources (46 in total) from the final statistical analysis. 
This leaves us with a sample of 112 sources. 
Of this sample of 112 sources, 10 were not detected in the NVSS catalogue and therefore we visually inspected the images as described above. For 6 sources there was a detection below 5$\sigma$ in the NVSS image.
Following this automatic procedure, we selected 19 sources as restarted candidates. 

As mentioned above, a final visual inspection was also carried out to prevent the inclusion of possible contaminants. Six sources turned out to have signatures of active jets fuelling the lobes and they were excluded because their high CP is most likely caused by Doppler boosting. This left us with 13 candidate restarted radio sources. 
Figure~\ref{fig:all_criteria1} shows the LOFAR 6$^{\prime\prime}$ and FIRST contours of the candidates selected using CP and low-SB (indicated by {\sl `CP'} in the upper right corner).

\begin{figure} [h]
\centering
  \includegraphics[width=7cm]{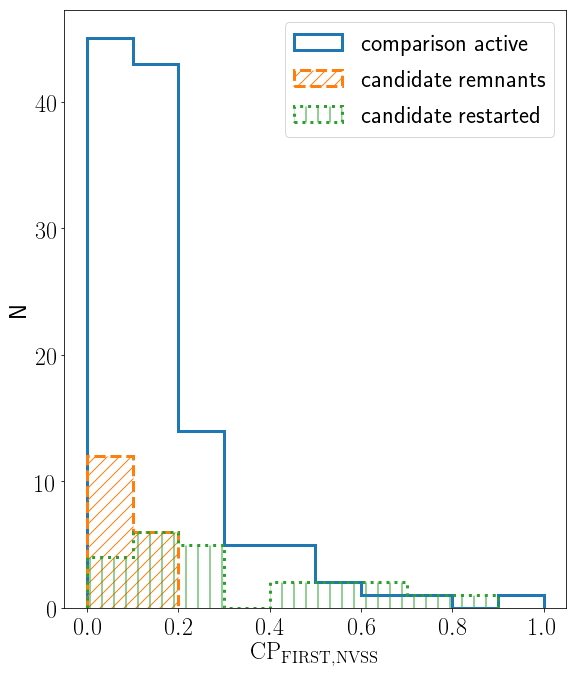}
\caption{Histogram showing the CP distribution of candidate restarted radio sources (green), candidate remnant radio sources (orange), where CP is mostly an upper limit as the majority of the radio cores are not detected in the FIRST images, and comparison active radio galaxies (blue). The restarted sample shown here includes all candidates selected with the different criteria, not only CP + low-SB; see Sect.~\ref{sec:results_restarted_selection}.}
\label{fig:allsample_cp1400_distr}
\end{figure}

The properties of the final 13 candidate restarted radio galaxies selected using the CP + low-SB criterion are listed in Table~\ref{tab:candidate_restarted} and Table~\ref{tab:candidate_restarted_appendix}.
The distribution of the CP for the candidate restarted radio sources is presented in Fig.~\ref{fig:allsample_cp1400_distr} in green. We would like to point out that in the plot we show not only sources selected with the CP + low-SB criterion as described here but also those selected with the criteria described in the following sections.

\subsubsection{Steep spectral index of the core}\label{subsec:results_SI}

Radio cores in radio galaxies at low redshifts are characterised by a flat or inverted spectral index that is due to self absorption (see e.g. \citealt{1979ApJ...232...34Blandford,1984A&A...139...55Feretti, 2008MNRAS.390..595Mullin}).
We note that due to the available resolution (as mentioned in Sect.~\ref{sec:scenarios_and_selection}), the computed spectral indices are expected to include a superposition of the actual nuclear activity and the jet basis or compact lobes. Therefore, observing a central compact region with a steeper spectral index may indicate the presence of newly restarted {\sl prominent} sub-arcsecond jets that cannot be resolved with the available data. We define a core as `steep spectrum' if the spectral index is $\alpha^{\rm 150 MHz}_{\rm 1400 MHz} \geq 0.7$ (e.g. \citealt{1979ApJ...231..299Readhead, 1988ApJ...334..552Marscher88}).
We chose this conservative value in order to take into account the large uncertainties in the flux density measurements that may affect the results (see below).
The spectral index of the core was calculated between 150 MHz and 1400 MHz. The core flux density measurement at 1400 MHz is described in the previous section. The flux density of the core from the LOFAR 150 MHz 6$^{\prime\prime}$ image was obtained in the same way (see Table~\ref{tab:candidate_restarted_appendix}). In particular, we measured the core flux density as the peak flux density at the location of the detection of the core at higher frequency (FIRST) or at the position of the optical counterpart; see Sect.~\ref{sec:optical_identification}

Extra care was required when deriving the spectral index of the cores using the flux density (or limit) at 1400 MHz and 150 MHz.
After applying the automatic criterion, we examined the FIRST and LOFAR 6$^{\prime\prime}$ images to check for possible contamination by extended emission seen in projection. This is more likely to occur at low frequencies where the core is often embedded in diffuse emission, which contaminates the flux density, making the spectrum of the core artificially steep. Automatic application of the spectral index criterion to our sample left us with 78 sources. However, visual inspection revealed that only in 6 out of 78 candidates is the core isolated in both FIRST and LOFAR 6$^{\prime\prime}$ images or isolated in the LOFAR 6$^{\prime\prime}$ image but not detected in FIRST. No detection in the FIRST image provides a lower limit to the spectral index, suggesting that sources with $\rm \alpha^{\rm 150 MHz}_{\rm 1400 MHz}>0.7$ are restarted candidates.
The error on the spectral index was computed using:
\begin{equation}
    \alpha_{err}=\frac{1}{\rm ln\frac{\nu_{1}}{\nu_{2}}}\sqrt{\left( \frac{S_{1, err}}{S_1}\right)^2+ \left( \frac{S_{2, err}}{S_2}\right)^2},
\end{equation}{}
where $S_1$ and $S_2$ are the flux densities at frequencies $\nu_1$ and $\nu_2$ and $S_{1,err}$ and $S_{2,err}$ are the respective errors. The error on the spectral index values is $\rm \pm$ 0.05, assuming a constant error of 11\% for the LOFAR measurements (Mandal et al. in prep.) and 5\% for those of FIRST \citep{1995ApJ...450..559BeckerFIRST}.

The LOFAR 6$^{\prime\prime}$ and FIRST contours of these six candidates selected based on their steep spectrum cores (SSC) overlaid on the LOFAR 18$^{\prime\prime}$ resolution maps are shown in Fig.~\ref{fig:all_criteria1}. These candidates are indicated with {\sl `SSC'} in the upper right corner, while the properties of the sources are listed in Table~\ref{tab:candidate_restarted} and Table~\ref{tab:candidate_restarted_appendix}.

\subsubsection{Visual inspection of the sample}\label{subsec:results_visual} 

We complemented the selection of restarted candidates with a visual inspection of the 158 sources in the sample. 
This was done to increase the sample completeness by adding sources showing for example extended inner jets surrounded by low-SB extended emission. One of the most obvious classes of sources belonging to this group is the DDRG and we identified three of them (J103621+564323, J104252+553536, J104809+573010). In particular, we classify DDRGs as those showing two clear peaks of emission along the total extent of the lobe in both lobes. In addition, we looked for sources with a bright inner region (not limited to the core) and low-SB extended emission using the LOFAR 150 MHz 18$^{\prime\prime}$ and 6$^{\prime\prime}$ images. In this way, we selected two more candidates (J102955+584621 and J110021+601630).

The LOFAR 6$^{\prime\prime}$ and FIRST contours of these five candidates (indicated with {\sl `V'} in the upper right corner) overlaid on the LOFAR 18$^{\prime\prime}$ resolution maps are shown in Fig.~\ref{fig:all_criteria1} and candidates with their properties are listed in Table~\ref{tab:candidate_restarted} and Table~\ref{tab:candidate_restarted_appendix}.

\begin{table*} [h]
\caption{List of candidate restarted radio galaxies selected with the different criteria described in Sect.~\ref{sec:results_restarted_selection}. The columns show: source name of the candidate restarted radio source; photometric ($^{\rm p}$) or spectroscopic ($^{\rm s}$) redshift from the SDSS DR14 or lower limit redshift from the WISE W1 magnitude; the flux densities at 150 MHz from the 18$^{\prime\prime}$ catalogue in mJy \citep{2016MNRAS.463.2997Mahony}; the radio CP computed as described in Sect.~\ref{subsec:results_CP}; SB at 150 MHz; spectral index of the inner region and the selection methods used to identify the source (V = visual identification, CP = high core prominence, SSC = steep spectral index of the core).}
\label{tab:candidate_restarted}
\centering
\begin{tabular} {l c l l l l l p{5cm}}
\hline\hline
Name & redshift & $\rm S_{int, 150MHz}$ & CP$_{\rm 1400 MHz}$   & $SB_{150MHz}$              & $\rm \alpha_{inner\:region}$& Selection  \\
                          &                      &      [mJy]                                         &          & [mJy $\rm arcmin^{-2}$]               & & criteria \\ \hline
  J103416+590523    &   0.47 $\pm$ 0.06 $^{\rm p}$   &   29.54   &   0.19    & 47    & -- & CP   \\
  J103508+583940    &   0.4709 $\pm$ 0.0001 $^{\rm s}$          &   34.17   &   0.80    & 46      & -- & CP       \\
  J103730+600011    &   0.02815 $\pm$ 0.00002 $^{\rm s}$        &   12.57   &   0.30    & 38      & -- & CP       \\
  J103841+563544    &   0.57 $\pm$ 0.07 $^{\rm p}$      &   44.04   &   0.70    & 50 & -- & CP       \\
  J104113+580755    &   0.30894 $\pm$ 0.00006 $^{\rm s}$        &   64.17   &   0.56    & 21      & -- & CP       \\
  J104204+573449    &   0.4808 $\pm$ 0.0001 $^{\rm s}$          &   34.81   &   0.26    & 21      & -- & CP       \\
  J104424+602917    &   0.22 $\pm$ 0.02 $^{\rm p}$   &   68.18   &   0.17    & 40    & -- & CP   \\
  J104520+563149    &   0.034 $\pm$ 0.1 $^{\rm p}$   &   35.42   &   0.20    & 22    & -- & CP   \\
  J104834+560005    &   0.796 $\pm$ 0.04 $^{\rm p}$     &   16.48   &   0.29    & 28 & -- & CP       \\
  J104912+575014    &   0.07256 $\pm$ 0.00002 $^{\rm s}$   &   127.26  &   0.60    & 45    & -- & CP   \\
  J105340+560950    &   > 0.365 &   36.80   &   0.42    & 40    & 0.72 $\pm$ 0.05 & CP, SSC  \\
  J105436+590901    &   0.8861 $\pm$ 0.0003 $^{\rm s}$   &   18.62   &   0.84    & 23    & -- & CP   \\
  J105524+561616    &   0.31 $\pm$ 0.03 $^{\rm p}$      &   26.93   &   0.29    & 39 & -- & CP       \\
  J103815+601111    &  0.19676 $\pm$ 0.00005 $^{\rm s}$    & 22.21     & -           & -     & > 0.80 $\pm$ 0.05 & SSC   \\
  J103845+594414    &   0.38 $\pm$ 0.03 $^{\rm p}$      &   42.65   &   0.49    & 38 & 0.80 $\pm$ 0.05 & SSC \\
  J105057+562349    &  0.62 $\pm$ 0.07 $^{\rm p}$     & 6.71      & -       & -      & > 0.76 $\pm$ 0.05 & SSC   \\
  J105418+595220    &  0.79 $\pm$ 0.04 $^{\rm p}$     & 24.65     & -       & -      & > 0.81 $\pm$ 0.05 & SSC   \\
  J105723+565938    &  > 1.1417      & 43.73     & > 0.34       & -         & 1.1 $\pm$ 0.05 & SSC       \\
  J102955+584621    & 0.39 $\pm$ 0.05 $^{\rm p}$  &  47.06    &  < 0.09 & 46      & -- &  V  \\
  J103621+564323    & 0.55 $\pm$ 0.05 $^{\rm p}$  & 579.19    &  < 0.01  & 184   & -- &  V; DDRG \\
  J104252+553536    & 0.5221 $\pm$ 0.0002 $^{\rm s}$  &  50.03    &  < 0.04  & 34   & -- &  V; DDRG \\
  J104809+573010        & 0.31742 $\pm$ 0.00007 $^{\rm s}$    &  38.82    &  0.15*   & 24        & -- &  V  \\
  J110021+601630    & 0.19857 $\pm$ 0.00003 $^{\rm s}$  &  294.30   &  < 0.06  & 49      & -- &  V  \\
\hline\hline
\end{tabular}
\end{table*}

\subsection{Final sample: description and caveats} \label{sec:implications_from_the_selection}
Following the procedure described above, we identified a sample of candidate restarted radio sources. Despite all the limitations described in Sect.~\ref{sec:results_restarted_selection}, we found up to 23 candidate restarted radio sources in our sample, corresponding to a fraction of at most 15\% of the starting sample of 158 sources.

In particular, we obtained 13 candidate restarted radio sources from the CP (combined with low-SB) selection and 6 from the SSC criterion. We note that one source was selected based on both criteria (see Table~\ref{tab:candidate_restarted}).
Thanks to the visual inspection of the sources in the sample we further identified three DDRGs (J103621+564323, J104252+553536 and J104809+573010) and two sources with a brighter inner region and low-SB extended emission (J102955+584621 and J110021+601630). 

The selection criteria applied may miss some cases of restarted candidates. For example, GPS sources would not be selected by the steep spectral index of the core criterion because they show an inverted spectral index when detected at low frequencies (\citealt{1998PASP..110..493Odea}). 
To investigate whether or not we missed some candidate restarted GPS radio sources, we checked the seven sources showing an inverted spectral index of the inner region between 150 and 1400 MHz. 
Only one radio source had a prominent inner region and a reliable measurement of the spectral index. This latter source was selected as a candidate restarted radio source based on the CP but did not satisfy the low-SB criterion and the image suggests the presence of the collimated features. It was therefore rejected from our final sample.

Another group of sources that might be missed by our selection consists of multi-component extended sources that were not automatically associated by the source extraction algorithm used to construct the catalogue (PyBDSF). This implies that there might be sources whose actual total angular size is > 60${^{\prime\prime}}$ that have been missed in our sample of 158 sources (and therefore in our restarted selection) because different source components have not been correctly associated. We note that this may be especially relevant for DDRGs, where the inner lobes can often be very well detached from the outer remnant lobes (see \citealt{2019A&A...622A..13Mahatma_restarted}). To overcome this issue, we further inspected the field looking for obvious cases of this nature but did not find any. We therefore conclude that the number of lost DDRGs must be zero or very low.

In particular, we missed those candidates where the old and new phases are only imprinted in the properties of the spectral index.
Therefore, sources such as 3C~388 (\citealt{1982ApJ...257..538Burns, 1994ApJ...421L..23Roettiger, 2020arXiv200313476Brienza388}) may be missed by the sample selection used above, although we know that some must be present in the LH field. For example, in the case of the visually selected candidate restarted radio source J104809+573010 shown in Fig.~\ref{fig:all_criteria1}, the two external lobes have much steeper spectral indices compared to the inner lobes, as already pointed out in the discussion of Appendix B of \citet{2016MNRAS.463.2997Mahony}. In the same paper, another candidate restarted radio source (J104912+575014) selected based on the CP (and low-SB) is presented. This source appears completely compact and unresolved even with deep Westerbork Synthesis Radio Telescope (WSRT) data (image presented in \citealt{2018MNRAS.481.4548Prandoni}) but shows extended emission at low frequency indicating a steep spectral index and therefore an old remnant structure. This source also shows an asymmetric morphology, which might be explained in a future more detailed environmental study (see e.g. \citealt{2019MNRAS.482.5625Rodman}).
A project for selecting restarted objects in the LH using the resolved spectral index properties is now in progress (Brienza et al., in prep).

\section{Results: properties of the host galaxy}\label{sec:derived_properties_of_the_host_galaxy}

In our sample, we have now identified the three groups of objects representing radio sources in different stages of their evolution: remnant candidates (using the results from \citealt{2017A&A...606A..98B}), restarted candidates from our selection described in Sect.~\ref{sec:results_restarted_selection}, and all the remaining radio sources considered to be the comparison sample of active radio galaxies.
In this section, we compare the host galaxy properties of all the sources in our sample, while in Sect.~\ref{subsec:discussion_radio_properties} we compare the radio properties of these three groups.

The optical/IR identification for the majority of the 158 sources in the sample (see Sect.~\ref{sec:optical_identification}) allowed us to derive stellar masses, as explained further in this section.
Comparing stellar masses and WISE colours of the host galaxies for the three groups of radio sources allows us to explore any differences that might be due to the effect that the repeated jet activity has on its host galaxy.

\subsection{Distribution of redshifts}\label{subsec:discussion_redshiftGroups}

\begin{figure} [h]
\centering
\minipage{\textwidth}
  \includegraphics[width=7.0cm]{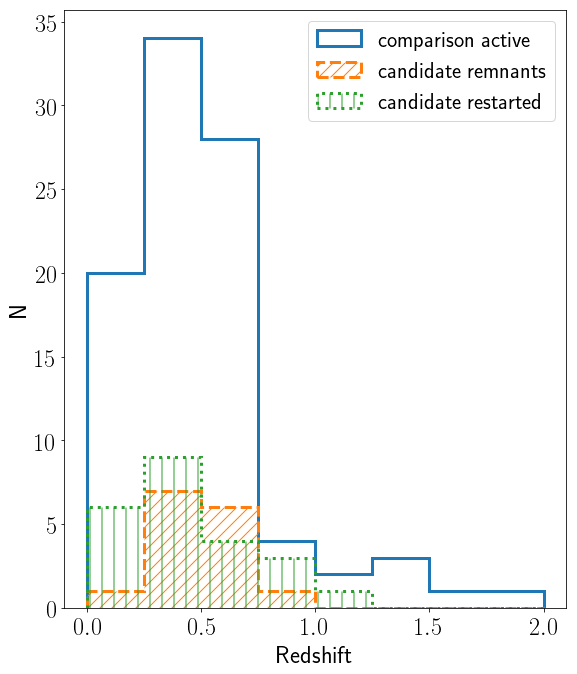}
\endminipage
\caption{Histogram showing redshift distribution of candidate restarted (green), remnant (orange), and active (blue) radio galaxies for the sources with an optical counterpart; see Sect.~\ref{sec:optical_identification} for details.}
\label{fig:allsample_redshift_distr}
\end{figure}

The distribution of redshifts is shown in Fig.~\ref{fig:allsample_redshift_distr} for the three subgroups of our sample.
Redshift values for the candidate restarted radio sources and their errors taken from SDSS DR14 or as lower limits from WISE (see Sect.~\ref{sec:optical_identification}) are reported in Table~\ref{tab:candidate_restarted}. As indicated in Table~\ref{tab:candidate_restarted}, 10 candidate restarted radio sources have a redshift obtained from the optical SDSS spectrum, 11 candidates have a photometric redshift, and 2 restarted candidates have a lower limit redshift value from WISE.
The median values of redshift for the three groups of radio sources and $p$-values from the Kolmogorov--Smirnov (KS) comparison of two datasets (a two-sided KS test; \citealt{1933Kolmogorov}) are reported in Table~\ref{tab:p_values}. The KS test was used to quantify the statistical difference between the distributions, at the 95\% confidence level. 
The ranges of the redshift distribution for the three groups are statistically similar. Therefore, we cannot reject the null hypothesis that these three samples come from the same distribution.

The lack of candidate remnants and restarted radio sources at high redshift that we see in Fig.~\ref{fig:allsample_redshift_distr} is not surprising. The diffuse emission is expected to fade rapidly at high redshift due to larger inverse Compton losses, and therefore by the time the radio galaxy is in the restarted phase, the remnant emission is more likely to have faded away \citep{2018MNRAS.475.2768Hardcastle}, meaning that these sources will be missed by our selection criteria. 

Our results are similar to those presented by \citet{2012ApJS..199...27Saripalli}. They present optical identifications for a large fraction (83\%) of their sample and about half of their restarting sample is found to be at a redshift lower than 0.5. Similarly, more
than half of our comparison (58\%), remnant (53\%), and restarted (65\%) samples are at z < 0.5.

Having identified the radio sources, we performed a first-order investigation of the environment properties.
The environment is an important parameter for understanding the physics of the propagation of the jets and the radio plasma evolution, and is a valuable input for the modelling (\citealt{2007A&A...470..875Parma2007,2011A&A...526A.148Murgia2011,2013MNRAS.430..174HardcastleKrause13,2015ApJ...806...59TurnerShabala, 2018MNRAS.480.5286Yates, 2019A&A...622A..12Hardcastle}, Shabala et al., accepted).

A number of studies suggest that remnant radio galaxies are more likely to reside in rich environments due to the presence of a dense intergalactic medium which is able to reduce the adiabatic expansion of the plasma and therefore prolong the visibility period of the remnant. However, this is not always the case (see e.g. B2~0924+30, \citealt{1987MNRAS.227..695Cordey,2017A&A...600A..65Shulevski17} and blob1, \citealt{2016A&A...585A..29Brienza}).

We used the NASA/IPAC Extragalactic Database (NED) automatic search for known clusters as an approximate environment analysis. We constrained the search in redshift and radius around the radio galaxy. The radius was constrained to 4${^{\prime}}$ for all of the sources, representing a radius of about 1500~kpc for the median redshift
of our sources. In this way, we found that 34 of the 158 sources (3 remnants, 4 restarted, and 27 comparison) in our sample reside in a single cluster according to NED classification. Only one cluster is a large Abell cluster (Abell 1132; \citealt{2018MNRAS.473.3536Wilber}) and two sources from the active comparison sample are part of it.

From this search, it does not appear that remnant and restarted radio galaxies in our sample are more likely to live in a dense environment than those of the parent population. This result is interesting but will need to be confirmed by a more thorough analysis in the future, when we have spectroscopic redshifts and a more complete crossmatch with the available catalogues of galaxy clusters.

\subsection{WISE colours}\label{subsec:discussion_WISE}
\begin{figure}
  \includegraphics[width=\linewidth]{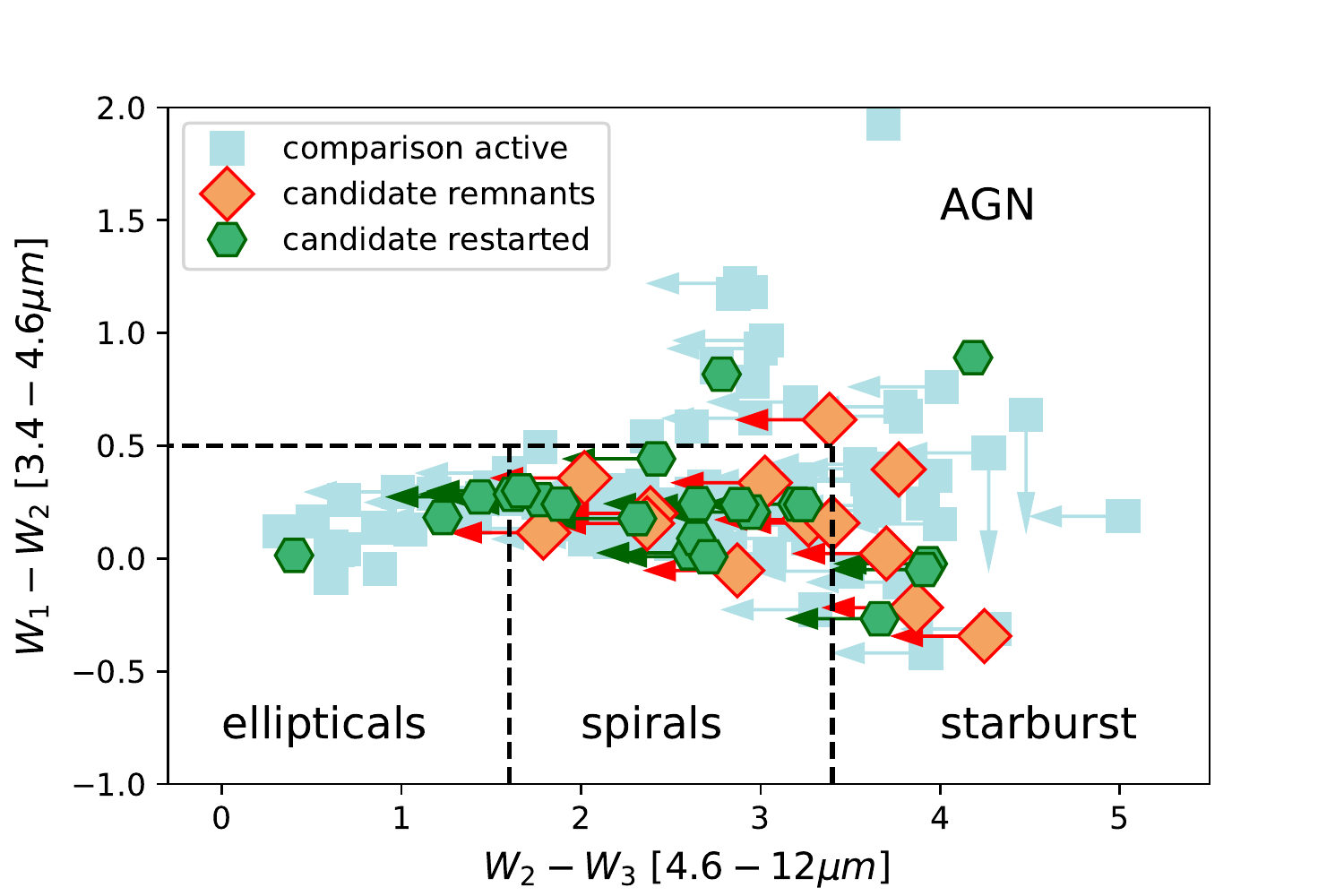}
\caption{WISE colour--colour diagram showing the sample of candidate remnant radio galaxies (orange diamonds), candidate restarted radio galaxies (green hexagons), and sample of active galaxies (blue squares). The upper limits are indicated by arrows. 
Classification follows \cite{2016MNRAS.462.2631Mingo}.}
\label{fig:WISE_colours}
\end{figure}

In order to characterise the properties of the host galaxies, we looked at their infrared colours and their distribution in the WISE colour--colour plot.
WISE short-wavelength bands (W1 and W2) primarily sample flux density from stellar photospheres. A higher value of W1 - W2 colour indicates dustier and/or increasingly star-forming objects, while a lower value indicates old stellar populations. 
Longer WISE wavelengths (W3 and W4) are more sensitive to warm dust emission heated by stars or of the dusty torus surrounding some accreting black holes (\citealt{2010AJ....140.1868Wright}). 
The WISE colour--colour plot is therefore used to classify the galaxies according to their dust content (e.g. \citealt{2010AJ....140.1868Wright, 2012ApJ...753...30Stern, 2013AJ....145....6Jarrett, 2016MNRAS.462.2631Mingo, 2017A&A...604A..43Maccagni}). The plot has also been used to classify their spectral properties (\citealt{2018MNRAS.480..707Prescott}), to compare host galaxy properties of different classes of radio sources (\citealt{2019A&A...622A..13Mahatma_restarted}), and to discriminate between star-forming galaxies and radio-loud AGN (\citealt{2019A&A...622A..12Hardcastle}). \citet{2014MNRAS.438.1149GurkanWISE} used the WISE data to diagnose accretion modes in radio-loud AGN.
From Fig.~\ref{fig:WISE_colours}, we can see that all three subclasses have similar distributions, indicating similar ages of stellar populations and dust content in the host galaxy. 
Candidate restarted radio galaxies do not occupy a special region in the WISE colour--colour space relative to the comparison active sample, indicating that both phases in the life cycle of radio galaxies are hosted by the same types of galaxies. The same result is reported in the study of \citet{2019A&A...622A..13Mahatma_restarted} on DDRGs.
These three samples are statistically similar (see Table~\ref{tab:p_values}) and are therefore statistically consistent with being drawn from the same population.

\subsection{Stellar masses}\label{subsec:discussion_stellar_mass_and_sfr}

Stellar mass ($\rm M_{\star}$) is a fundamental property required for the description of galaxy evolution. There are many different ways to derive stellar masses, but for the purpose of this paper and given the available data, we used WISE short-wavelength bands. Another interesting property to estimate is the SFR but due to the strong AGN contribution at 12 $\mu$m, it was not possible to calculate SFRs based on the infrared data we had.

\begin{figure} [h]
\centering
  \includegraphics[width=7.1cm]{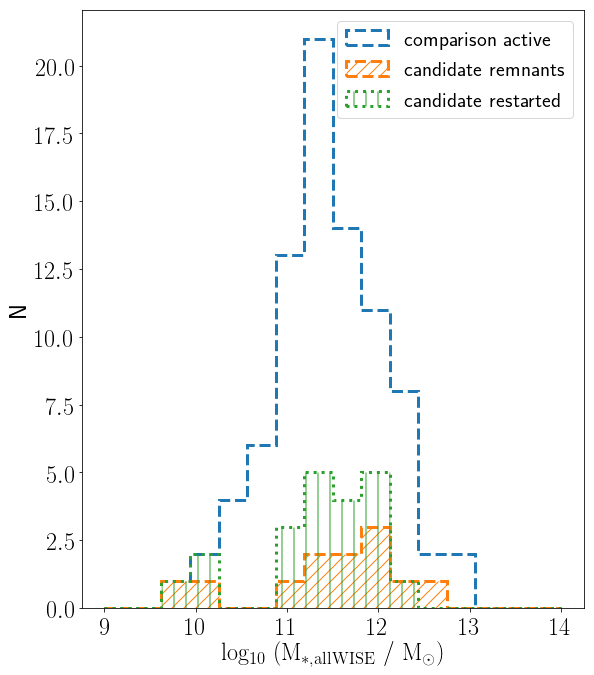}
\caption{Histogram showing stellar masses of candidate restarted (green), remnant (orange), and comparison active (blue) radio galaxies, in solar masses.}
\label{fig:histogram_stellar_mass_sfr}
\end{figure}

Infrared emission at 3.4 $\rm \mu$m and 4.6 $\rm \mu$m from galaxies mainly traces the old star population and has been shown to be an effective measure of galaxy stellar mass (\citealt{2013AJ....145....6Jarrett, 2014ApJ...788..144Meidt, 2014ApJ...782...90Cluver}).
Figure~\ref{fig:WISE_colours} shows that W2 is not strongly AGN dominated. We therefore used the formula from \citet{2014ApJ...782...90Cluver}:
\begin{equation}
{\rm log_{10}}M_{\star}/L_{3.4} = -1.96 \times (W1 - W2) - 0.03
,\end{equation}
where $\rm L_{3.4}(L_{\odot}) = 10^{-0.4(M-M_{\odot})}$ and M is the absolute magnitude of the source at 3.4 $\rm \mu$m and $\rm M_{\odot}$ = 3.24 (\citealt{2013AJ....145....6Jarrett}).
From the sources in our sample that were detected in WISE and had an optical counterpart, we removed six broad-line AGN based on the available SDSS spectra because their optical and IR flux densities are likely dominated by emission from the central AGN. This step left us with 117 sources for the calculation of the stellar masses.
The distribution of stellar masses of the galaxies is shown in Fig.~\ref{fig:histogram_stellar_mass_sfr} and ranges between $10^9$ and $10^{12}$ M$_\odot$, consistent with the stellar masses of massive elliptical galaxies. 
The high-mass (M > $\rm 10^{12}$ M$_\odot$) tail of the distribution is likely the result of some contamination by the AGN, that is, the IR emission may be contaminated by AGN-heated dust and therefore stellar masses are overestimated. 

The values of stellar masses and their respective errors of 21 candidate restarted radio sources are listed in Table~\ref{tab:candidate_restarted_appendix}. For the error estimate, we propagated errors from redshift and WISE colours.
The median values of stellar masses and $p$-values from the KS test of the three samples shown in Fig.~\ref{fig:histogram_stellar_mass_sfr} are listed in Table~\ref{tab:p_values}.
According to the two-sided KS test, we cannot reject the null hypothesis that these three samples come from the same distribution.

We conclude that there is no difference in the stellar masses of the host galaxies of the three groups, which implies that we are looking at similar types of galaxies hosting radio sources in a different phase of their evolution. 

\begin{figure*} [!h]
\centering
  \includegraphics[width=7cm]{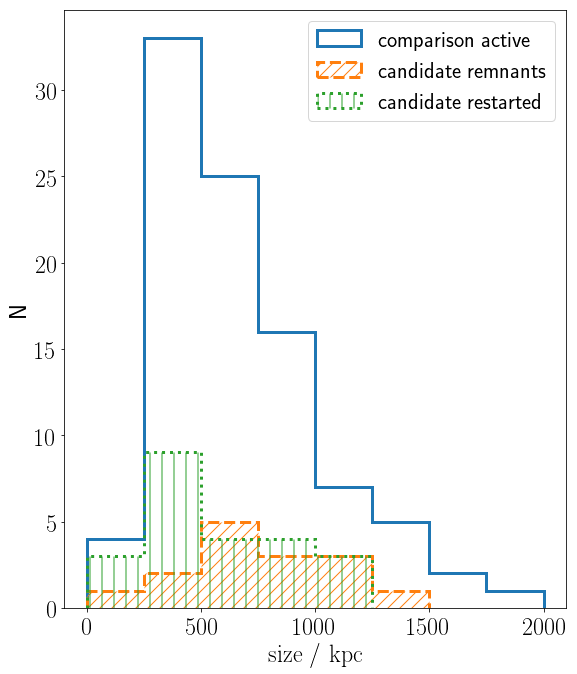}
  \includegraphics[width=7cm]{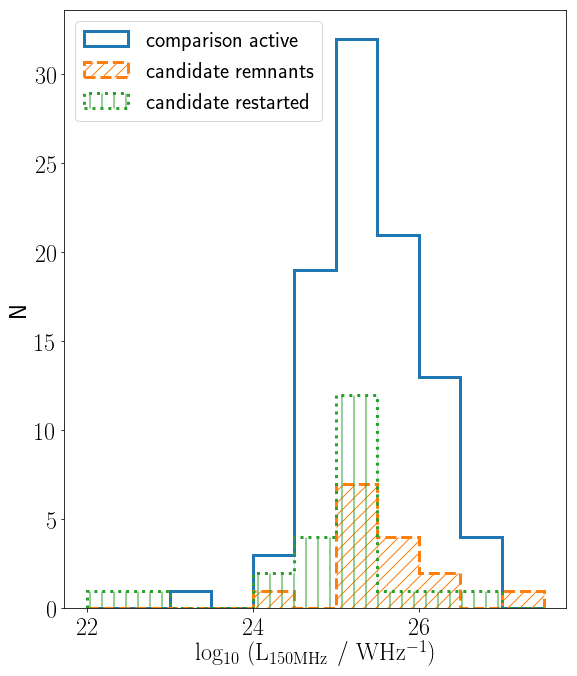}
\caption{Histogram showing linear sizes in kiloparsec (left) and radio luminosity in WHz$\rm ^{-1}$ at 150~MHz (right) of candidate restarted (green), candidate remnant (orange), and comparison active (blue) radio galaxies with optical identification from the SDSS or lower limit in redshift from WISE (see text for details).}
\label{fig:allsample_size_power150_distr}
\end{figure*}

\begin{figure} [ht]
\centering
  \includegraphics[width=9.5cm]{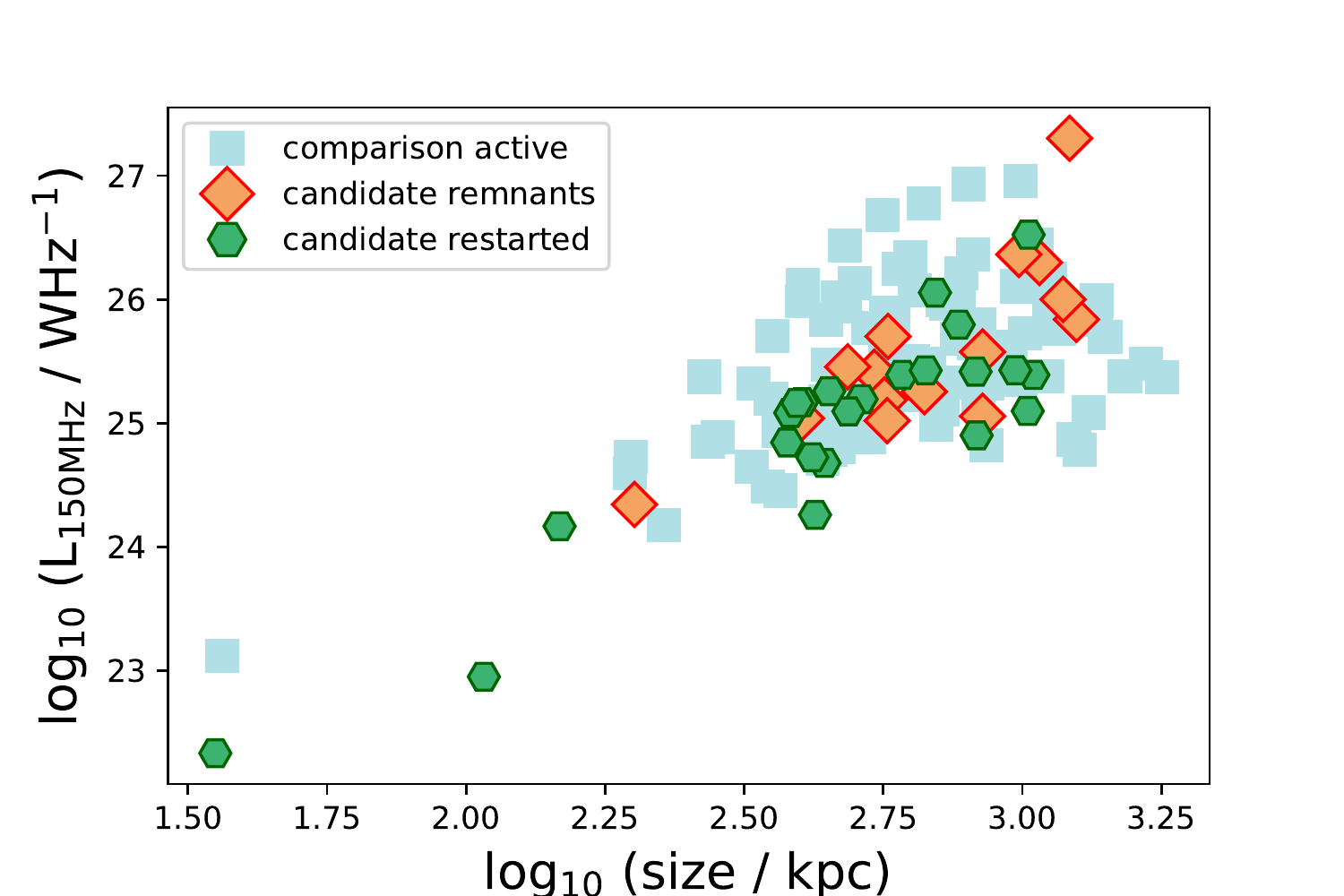}
\caption{Radio luminosity--linear size diagram for the sources in our sample with the optical identification in SDSS. The radio luminosity $L_{150}$ is defined as the 150 MHz luminosity and the linear size in kiloparsecs is the physical extent of the radio source inside of its 3$\sigma_{local}$ contours at 150 MHz.}
\label{fig:power_size_diagram}
\end{figure}

\section{Radio properties: linear sizes and radio luminosities}\label{subsec:discussion_radio_properties}
The angular sizes of the radio sources in the full sample were measured along the maximum extent of the source using the 3$\sigma_{\rm local}$ contours of the 18$^{\prime\prime}$ image as a reference. 
The histograms of the distributions of the linear sizes for the three groups of sources are shown on the left-hand side of Fig.~\ref{fig:allsample_size_power150_distr}. This histogram includes the values of all sources in our sample with optical/IR identifications and therefore redshift values.
Linear sizes of all remnant candidates are given in Table~\ref{tab:candidate_restarted_appendix} together with their uncertainties propagated from redshift uncertainties.
The median values of the sizes and $p$-values of the three groups are reported in Table~\ref{tab:p_values}. 

Looking at the median values, the size of the candidate remnant and active comparison radio sources seem to be larger than those of the candidate restarted sample.
However, remnant candidates and active radio sources are at a higher median redshift than candidate restarted radio sources which could explain the discrepancy in size. Based on the $p$-values referring to the distributions of our three samples, we cannot reject the null hypothesis that these three samples come from the same distribution.

In addition to the size, we can also compare the radio luminosity of the three groups of objects. 
The distribution of the radio luminosity is shown in Fig.~\ref{fig:allsample_size_power150_distr} (right) for the sources with optical/IR identifications. The values of radio luminosities of 23 candidate restarted radio sources are listed in Table~\ref{tab:candidate_restarted_appendix}. Median values of the radio luminosities and $p$-values from the KS test are listed in Table~\ref{tab:p_values}. In addition, we performed a Wilcoxon-Mann-Whitney test \citep{2016Marx} to check the significance of the median luminosity differences, which gave a $p$-value of 0.02 for the restarted candidates and the active comparison sample and a $p$-value of 0.51 for the remnant candidates and the active comparison sample. 
Due to the different redshift distributions of the sources in our sample (see Fig.~\ref{fig:allsample_redshift_distr}), we decided to repeat the KS test for the redshift-matched samples. We took into account only sources between redshift 0.3 and 0.5, where our sample is complete (see Table~\ref{tab:p_values}).
Therefore, we cannot reject the null hypothesis that the three samples come from the same distribution. Radio luminosities and respective uncertainties of all restarted candidates are given in Table~\ref{tab:candidate_restarted_appendix}. To calculate the errors, we propagated uncertainties in redshift and flux densities. For the latter, we take into account uncertainty reported in the PyBDSF catalogue and uncertainty in the flux density scale - 11\%
(Shimwell et al. 2019, Mandal et al. in prep). Uncertainties in
measured source size are less than 10\%\ for all sources, while
luminosity uncertainties are less than 20\%\ (see Table~\ref{tab:candidate_restarted_appendix}).
Hence the distributions plotted in Fig.~\ref{fig:allsample_size_power150_distr} are robust.

Figure~\ref{fig:power_size_diagram} shows a luminosity--size diagram (i.e. the linear size versus the 150~MHz radio luminosity) of the candidate restarted, remnant, and comparison samples. This diagram is often used to trace the evolution of the radio properties in the radio-loud AGN population. 
From this plot, we can see that all three samples have similar ranges of both size and radio luminosity.

\begin{table*} [!h]
\caption{Table showing median values of optical and radio properties and $p$-values of the KS tests, where AC is the active comparison sample, cREMN are the candidate remnant radio galaxies, and cREST are the candidate restarted radio galaxies; $^{z}$ denotes statistics done on redshift-limited samples.}
\label{tab:p_values} 
\resizebox{\linewidth}{!}{
\centering
\begin{tabular}{l c c c c c c c}
\hline\hline
\textbf{property}   & $\rm AC_{median}$   & $\rm cREMN_{median}$    & $\rm cREST_{median}$    & $p$-value     & $p$-value     & $p$-value$^{z}$     & $p$-value$^{z}$ \\
                    &               &                   &                   & AC - cREMN    & AC - cREST    & AC - cREMN    & AC - cREST \\ \hline
redshift            & 0.463         & 0.486             & 0.394             & 0.670         & 0.726         & -             & -  \\ \hline
size/kpc            & 568.568       & 666.926           & 486.961           & 0.246         & 0.566         & 0.32          & 0.93  \\ \hline
radio luminosity/W Hz$^{-1}$   & 25.377        & 25.455            & 25.164            & 0.683         & 0.015         & 0.62          & 0.12  \\ \hline
$M_{\star}$/$M_{\sun}$ & 2.8 $\times$ $10^{11}$ & 6.2 $\times$ $10^{11}$ & 3.1 $\times$ $10^{11}$  & 0.259 & 0.986     & -             & -  \\ \hline
W1-W2 & 0.24 & 0.16 & 0.24 & 0.20 & 0.28 & - & - \\ \hline
W2-W3 & 2.62 & 3.27 & 2.64 & 0.20 & 1 & - & - \\ \hline \hline
\end{tabular}}
\end{table*}

\section{Discussion}\label{sec:discussion}
In this study, we were able to derive a sample of radio galaxies, which includes sources in different phases of the AGN life cycle. 
With the combination of spectral and morphological criteria, we systematically selected a subsample of candidate restarted radio sources from a parent sample of 158 radio sources larger than 60$^{\prime\prime}$.
We complemented this with the remnant radio sources already presented by \citet{2017A&A...606A..98B}. 

Thanks to the variety of criteria applied for the selection, restarted radio galaxies selected in our sample cover a wide range of morphologies going beyond the classical DDRGs.
Although it may not provide a complete census of the properties of restarted radio sources, our selection represents the first attempt to obtain, in a mostly automatic way, restarted radio galaxies of different morphologies and properties based on a collection of criteria already used in the literature.
For example, the visual CP and low-SB criterion was used by \citet{2012ApJS..199...27Saripalli} for the selection of candidate remnant and restarted radio galaxies. The SSC criterion was used in a number of studies including \citet{1981AJ.....86.1294Bridle} and \citet{1981ESASP.162..107Schilizzi} when studying the well-known restarted radio sources 3C~293 and 3C~236. Also see \citealt{2018MNRAS.473.4926Sebastian} for a more recent example.\\ 
A final visual inspection was required for the restarted candidates selected using CP + low-SB criterion in order to reject sources possibly affected by beaming.
In order to derive the SSC we had to pay attention to the contamination of the core flux density by the extended diffuse emission at low frequencies. Because of this, the number of restarted candidates selected based on the steep spectrum core likely represents a lower limit.

\subsection{Morphology of candidate restarted radio galaxies} \label{subsec:morphology_discussion}
We find a number of candidate restarted radio galaxies with interesting radio morphologies, often reminiscent of known objects. 
\citet{2012ApJS..199...27Saripalli} report pronounced asymmetry in the lobe extents of both restarting FRI and FRII sources, which we also notice in our sample. In particular, sources J104912+575014 and J102955+584621 extend in only one direction with respect to the core. This morphology might be the result of a difference in the density in the southern and northern lobes or could be due to the inclination angle where the northern lobe has already faded below the detection limit.
As already mentioned in Sect.~\ref{sec:implications_from_the_selection}, the southern lobe of J104912+575014 has not been detected in deep 1.4 GHz observations (\citealt{2018MNRAS.481.4548Prandoni, 2016MNRAS.463.2997Mahony}; Fig.~\ref{fig:all_criteria1}) but is detected only at low frequencies, implying that it has an ultra-steep spectrum ($\alpha^{1400}_{150}$ > 1.2). The morphology and the spectral property of the southern lobe appear to suggest that it belongs to the group of older restarted radio sources in our sample. 
There are also two restarted candidates (J103416+590523 and J104424+602917) in which emission has been observed in the direction perpendicular to the main lobe. This might be the indication of a change of the jet orientation between the two episodes of activity, such as the X-shaped radio source J0009+1244 (4C~12.03) from \citet{2017MNRAS.471.3806Kuzmicz}; Markarian 6 studied by \citet{2011ApJ...731...21Mingo_Markarian6} and \citet{2014MNRAS.440.2976Kharb_Markarian6}; or the change of a radio galaxy to a blazar (PBC~J2333.9-2343) studied by \citet{2017A&A...603A.131Hernandez}. It is also possible that the extended emission perpendicular to the jets axis is due to anisotropic pressure gradients in the hot atmospheres pushing the radio lobes in that direction, as can be seen in the buoyant backflow and overpressured cocoon models (see e.g. \citealt{1984MNRAS.210..929LeahyWilliams, 2002A&A...394...39Capetti,2012ApJ...746..167HodgesKluckReynolds}). Recently, \citet{2019MNRAS.488.3416HardcastleNGC326} proposed that the X-shaped radio structure observed in NGC~326 at low frequency is the result of the complex large-scale bulk motions within the X-ray-emitting medium induced by the ongoing cluster merger coupled with motions of the host galaxy itself with respect to that medium.
Finally, there are two DDRGs (J103621+564323 and J104252+553536), which form the least dominant group of restarted radio galaxies, in agreement with what was found by \citet{2012ApJS..199...27Saripalli}. On the other hand, \citet{2017MNRAS.471.3806Kuzmicz} report that the majority (i.e. 85\%) of the sources in their sample are classified as DDRGs. However, this
mostly reflects the selection of their sample (see Sect.~\ref{sec:introduction}) making a direct comparison impossible.
One of the DDRGs in our sample, J103621+564323, also shows hotspots. While this has been observed for the FRII restarted radio galaxies by \citet{2012ApJS..199...27Saripalli} and for the DDRGs by \citet{2019A&A...622A..13Mahatma_restarted}, it is instead uncommon for the rest of the candidate restarted radio galaxies in our sample. 

\subsection{Occurrence and implications for the life-cycle}\label{sec:discussion_occurence}
The goal of this paper is to derive the fraction of candidate restarted radio sources and compare this with the fractions of candidate remnant radio sources and the comparison sample of currently active radio galaxies, in order to gather information about the candidate restarted radio sources in terms of the duty cycle.
We derived the fraction of candidate restarted radio sources (having linear sizes larger than $60^{\prime\prime}$) to be up to 15\%.\ 
This can be compared with the upper limit of 11\% of candidate remnant radio sources derived for our sample.

It is worth noting that the fraction of candidate remnant radio sources in our sample is consistent with other studies and in particular with the 9\% of remnants found by \citet{2018MNRAS.475.4557Mahatma_remnants}. A lower fraction of remnants (only about 3\%) was found by \citet{2012ApJS..199...27Saripalli} at higher frequencies.
Thus, the fraction of restarted candidates appears to be (slightly) higher than or consistent with the fraction of remnants.
This is interesting because it may indicate the presence of a rapid duty-cycle in these radio sources.
Our result suggests that in a large number of remnant radio galaxies, the activity restarts before the remnant emission fades away.

\citet{2017A&A...606A..98B}, \citet{2017MNRAS.471..891Godfrey} and \citet{2018MNRAS.475.2768Hardcastle} presented modelling of the remnant population assuming radiative and dynamical evolution models. All three of these latter studies conclude that remnant sources fade rapidly.

The modelling presented in \citet{2017A&A...606A..98B} and \citet{2017MNRAS.471..891Godfrey} further suggests that most of the observed remnant radio galaxies are relatively young (with total ages between 5 $\times$ $10^7$ and $10^8$ yr; see Fig. 6 in \citealt{2017A&A...606A..98B}). 
As discussed in \citet{2017A&A...606A..98B}, the modelling suggests that about 70\% of the remnants have ages < 1.5 $\times$ $t_{\rm on}$. This would imply a $t_{\rm off}$ < $t_{\rm on}$/2.
Therefore, the majority of the remnant sources would be observed soon after the switch-off of the radio source and they are expected to evolve quickly due to dynamic expansion.

Considering the aforementioned timescales for remnant sources and based on the fact that the candidate restarted sources that we have selected started their second phase of activity when the remnant lobes were still observable we can infer that the inactive phase of our restarting candidates should last a few tens of millions of years. All this implies that radio galaxies restarting on timescales longer than a few tens of millions of years or more will remain difficult to identify.

We stress that the age values discussed here should be taken with care as they are based on simulations which rely on a number of assumptions. One of these is the magnetic field which is typically computed using the often unrealistic equipartition conditions. For example, a factor two difference in the magnetic field value translates directly into about a factor two age difference. Another important parameter is the assumed total source age distribution which is assumed to be different in the various simulations.
For example, in the case of the modelling presented in \citet{2017A&A...606A..98B} and \citet{2017MNRAS.471..891Godfrey}, the distribution of $t_{\rm on}$ was taken to be a truncated log-normal distribution between 20 and 200 Myr, with a median of 30 Myr. 
\citet{2018MNRAS.475.2768Hardcastle} on the other hand explored two types of models, with ages being either uniformly distributed between 0 and 1000 Myr, or linearly distributed in log space between 1 and 1000 Myr. These latter authors found that the uniform-age models efficiently explain the size and luminosity statistics of bright sources observed by LOFAR, while the log-uniform models were a better representation of the faint population. Similarly, Shabala et al. (accepted) showed that power-law age-distribution models (i.e. models where the radio-source population is dominated by short-lived sources) can explain the observed properties of both active and remnant and/or restarted LOFAR sources. These findings are consistent with the modelling assumptions of \citet{2017A&A...606A..98B} and \citet{2017MNRAS.471..891Godfrey}, as the majority of short-lived sources will simply never grow large or luminous enough to make it into the observable LOFAR sample\footnote{Shabala et al. (accepted) recover the age distribution of {\it detected} active LOFAR sources to be log-normal, with a standard deviation of $\sim 0.2$ dex, albeit with older median ages by $\sim 0.5$ dex compared to \citet{2017A&A...606A..98B} and \citet{2017MNRAS.471..891Godfrey}.}. Modelling by both Shabala et al. (accepted) and \citet{2018MNRAS.475.2768Hardcastle} confirms the picture in which the remnant lobes fade quickly below the LOFAR detection limit. This fading is more rapid for older sources. 
Future studies, including modelling of the spectral index and follow-up observations, will help us to put tighter constraints on the age of the candidate restarted radio galaxies and the time that passed between the two bursts of activity.

Our results are also consistent with \citet{2019A&A...622A..17Sabater}.
These latter authors conclude that the most massive galaxies ($> \rm 10^{11}$ $M_{\sun}$, consistent with our galaxies) are always switched on when radio luminosities down to log ($L_{\rm 150~MHz}/\rm W Hz^{-1}$) = {21.7} are considered. The sources in our sample have radio luminosities higher than those of \citet{2019A&A...622A..17Sabater}, with the median value of log ($L_{\rm 150~MHz}/\rm W Hz^{-1}$) = 25.164. Therefore, although we do not expect to find such an extremely rapid duty cycle, we suggest that the high fraction of restarted sources is in line with the results of \citet{2019A&A...622A..17Sabater} for the sources of that mass and luminosity.

\subsection{The parent population: comparing radio and optical properties}

Another result from the present study is that, based on the radio and optical properties of our sample, we conclude that there is no statistical difference in the host galaxies of our three samples of candidate remnant radio galaxies, candidate restarted radio galaxies, and the active comparison sample. We do not find any indication of a correlation between a phase in the duty cycle with the environment or host galaxy.

If, as described above, restarted activity appears relatively soon after the remnant phase, 
we would expect remnant radio galaxies to be similar in linear size to restarted radio galaxies. Results from this paper confirm this. Indeed, we do not see any statistically significant difference between the three phases in our sample of radio galaxies. 
This also supports the findings of \citet{2017A&A...606A..98B, 2017MNRAS.471..891Godfrey, 2018MNRAS.475.2768Hardcastle} and Shabala et al., (accepted) that the evolution of remnant plasma happens on short timescales.
Contrary to our results, \citet{2012ApJS..199...27Saripalli} found that their candidate restarted radio sources are on average 100 kpc larger than those in their comparison active sample.
In broad agreement with our results, the candidate restarted (mostly DDRGs) radio sources in the sample from \citet{2017MNRAS.471.3806Kuzmicz} cover a wide range of linear sizes from 0.02 to 876 kpc for the inner lobes to 48 to 4248 kpc for the outer lobes. 

Finally, the environment plays a role in the propagation of jets and possibly in the detectability of the remnant emission. However, we find that the majority of our candidate remnant and restarted radio sources do not reside in rich clusters and dense medium, though a more thorough study is needed.
Four restarted candidates reside in a denser environment or a cluster: J110021+601630, J102955+584621, J104912+575014 and J104424+602917, two of which are selected based on the CP and the other two visually. It is interesting to see from our radio images that the sources with larger asymmetries (see Sect.~\ref{subsec:morphology_discussion}) are in this group, possibly showing asymmetric morphology as a result of being influenced by the surrounding cluster environment.
The fact that the majority of candidate restarted radio sources do not reside in a dense medium supports the idea that the remnant and restarted phase that we observe happen on short timescales since we are able to observe both old and new radio emission. Both the recent LOFAR results on radio source lifetimes (\citealt{2019A&A...622A..12Hardcastle}) and earlier studies (e.g. \citealt{2008MNRAS.388..625Shabala}) show that compact, low-luminosity AGN dominate volume-limited low-redshift samples. If this can be interpreted in terms of the duty cycle, most jets must switch on and off on timescales of a few tens of millions of years. The companion paper (Shabala et al., accepted) also confirms this picture.
Characterisation of the environment will be key for the future studies of larger samples.

In conclusion, our study suggests that all three groups come from the same parent population, allowing us to use the active comparison sample as the initial sample for the modelling of the candidate remnant and restarted radio sources. In a companion paper (Shabala et al., accepted), we show that self-consistent modelling of active, remnant, and restarted sources can place important constraints on the radio-source duty cycle, and ultimately the plausible mechanisms responsible for the re-triggering of jet activity. Discussion about feedback and gas properties is deferred to future work.

\section{Summary and conclusions}\label{sec:summary_and_conclusions}
We selected a sample of 158 radio galaxies with size $> 60^{\prime\prime}$ from the LH area covered by LOFAR observations at 150 MHz with a resolution of 18$^{\prime\prime}$ and 6$^{\prime\prime}$.
From this sample, a group of remnant candidates -- corresponding to up to 11\% of the
objects -- were already selected by \citet{2017A&A...606A..98B}. To complement
this and shed light on the life cycle of radio sources, we selected candidate restarted radio galaxies by applying three different criteria: the CP combined with low-SB of the extended emission; the steep spectrum of the inner region; and a
visual inspection. The remaining objects in the sample provided a comparison sample for our analysis.

We find between 13 and 15\% restarted candidates
depending on the CP cut and the error in the spectral index. The (slightly) larger fraction, compared to candidate remnant radio sources, suggests that the restarted phase can often start after a relatively short remnant phase, i.e. resulting in the diffuse low-SB structure being still visible.
The fraction of restarted radio galaxies found in this work is lower than what was found by \citet{2012ApJS..199...27Saripalli}.

Thanks to the optical identification of the sample, we were able to compare the radio and optical properties of the different groups of objects. We report no difference in the radio and optical properties, extending the result obtained for DDRGs in the paper by \citet{2019A&A...622A..13Mahatma_restarted} to candidate restarted radio galaxies showing a broad variety of radio morphologies.
This is different from what is found by \citet{2017MNRAS.471.3806Kuzmicz}, as discussed in \citet{2019A&A...622A..13Mahatma_restarted}. 
The similarities between the host galaxies of the radio sources in different phases of the life cycle suggest that the remnant and restarting phase are not a consequence of changes in the host galaxy and that these two phases come from the same parent population.
These similarities also suggest that the restarting activity does not occur preferentially in a specific class of host galaxy or environment.
The results presented here expand our knowledge of the restarted radio sources beyond the well-known class of DDRGs.

The sample presented here and the classification of remnants and restarted radio galaxies are now providing the input for a modelling analysis of the evolution of the radio source aimed at interpreting and describing our results. This modelling, presented in Shabala et al., (accepted), is based on the Radio AGN In Semi-Analytic Environments (RAiSE) model (\citealt{2015ApJ...806...59TurnerShabala, 2018MNRAS.473.4179TurnerRAISEII}; \citealt{2018MNRAS.474.3361TurnerRAISEIII}) and provides an expansion to the results on the remnants presented by \citet{2017MNRAS.471..891Godfrey} and \citet{2017A&A...606A..98B}. 

The results presented here were derived from LOFAR observations of a relatively small region of sky (the LH area; $\sim$ 30 deg$^{\rm 2}$). A much larger area is now covered by the 
LoTSS survey (e.g. 400 deg$^{\rm 2}$ in the HETDEX area presented by \citealt{2019A&A...622A...1Shimwell}) and will allow for a major expansion of the work presented here. It is worth noting the importance of deep, high-resolution radio observations and good ancillary data for the study of the life cycle of radio sources.
In order to confirm the restarted nature of our candidates, follow-up observations are needed. For the observations of the inner regions, we plan to follow up candidate restarted radio sources with observations at a higher frequency and higher resolution. For the study of the remnant emission coming from the previous cycle of activity, observations at a higher frequency with the same resolution should be obtained in order to construct spectral index maps and investigate spectral curvature. 

\begin{acknowledgements}
LOFAR, the Low Frequency Array designed and constructed by ASTRON (Netherlands Institute for Radio Astronomy), has facilities in several countries, that are owned by various parties (each with their own funding sources), and that are collectively operated by the International LOFAR Telescope (ILT) foundation under a joint scientific policy.
This publication makes use of data products from the Wide-field Infrared Survey Explorer, which is a joint project of the University of California, Los Angeles, and the Jet Propulsion Laboratory/California Institute of Technology, and NEOWISE, which is a project of the Jet Propulsion Laboratory/California Institute of Technology. WISE and NEOWISE are funded by the National Aeronautics and Space Administration.
This research has made use of the NASA/IPAC Extragalactic Database (NED), which is operated by the Jet Propulsion Laboratory, California Institute of Technology, under contract with the National Aeronautics and Space Administration. This research made use of APLpy, an open-source plotting package for Python hosted at http://aplpy.github.com.
MB acknowledges support from the ERC-Stg DRANOEL, no 714245.
MB and IP acknowledge support from INAF under the PRIN SKA/CTA project ‘FORECaST’. IP acknowledges support from INAF under the PRIN MAIN STREAM project “SAuROS”. VHM thanks the University of Hertfordshire for a research studentship [ST/N504105/1]. BM acknowledges support from the UK Science and Technology Facilities Council (STFC) under grants ST/R00109X/1 and ST/R000794/1. PNB is grateful for support from the UK STFC via grant ST/R000972/1. JS is grateful for support from the UK STFC via grant ST/R000972/1.
SS thanks the Australian Government for an Endeavour Fellowship, 6719\_2018. SS is grateful to the Centre for Astrophysics Research at the University of Hertfordshire, and the Netherlands Institute for Radio Astronomy (ASTRON), for their hospitality.
\end{acknowledgements}

\bibliographystyle{aa} 


\appendix
\section{Figures of candidate restarted radio galaxies}
Here we show all the candidate restarted radio galaxies.
\begin{figure*}
    \minipage{\textwidth}
        \includegraphics[width=0.24\linewidth]
        {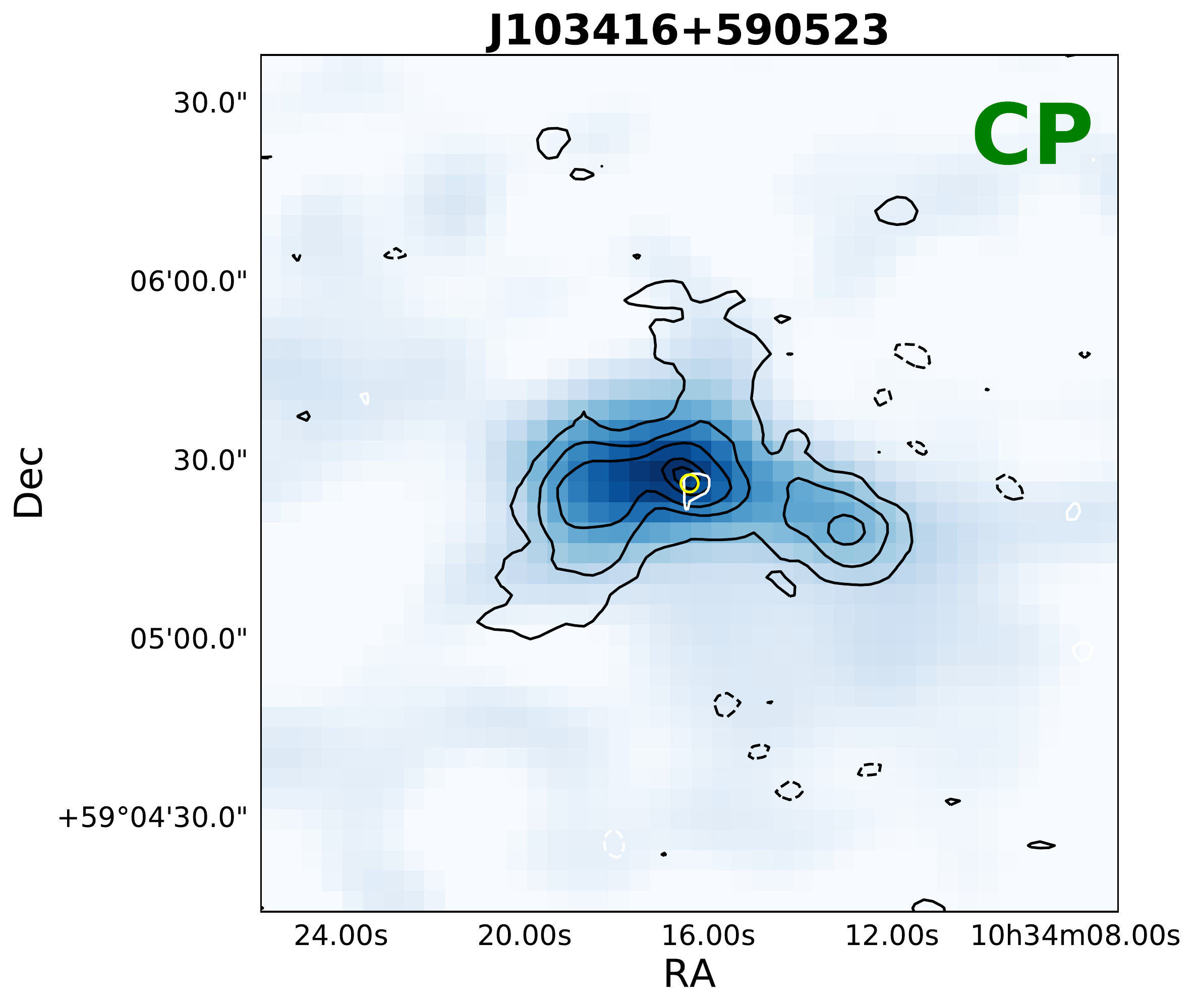}  
        \includegraphics[width=0.24\linewidth]
        {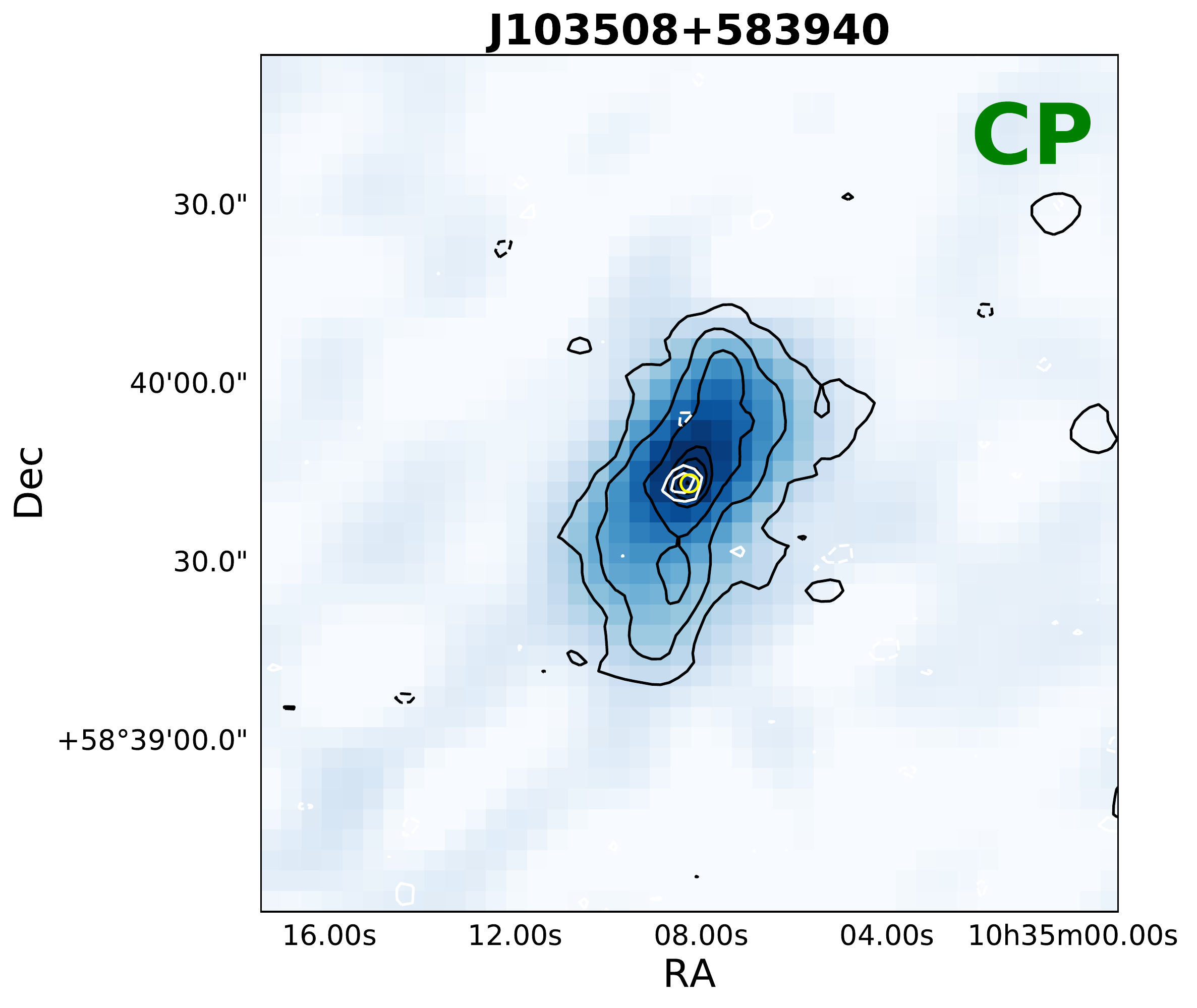}
        \includegraphics[width=0.24\linewidth]
        {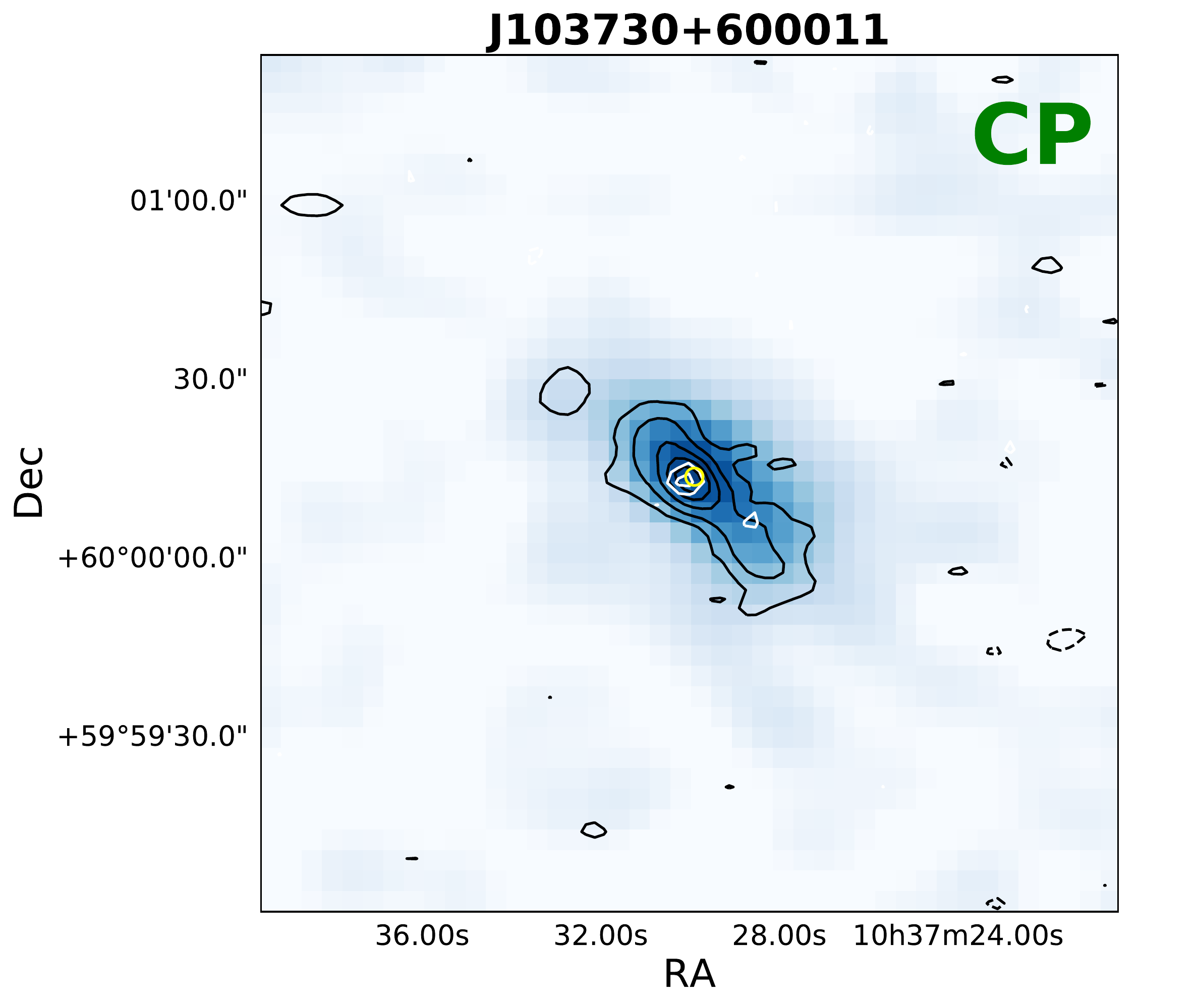} 
        \includegraphics[width=0.24\linewidth] 
            {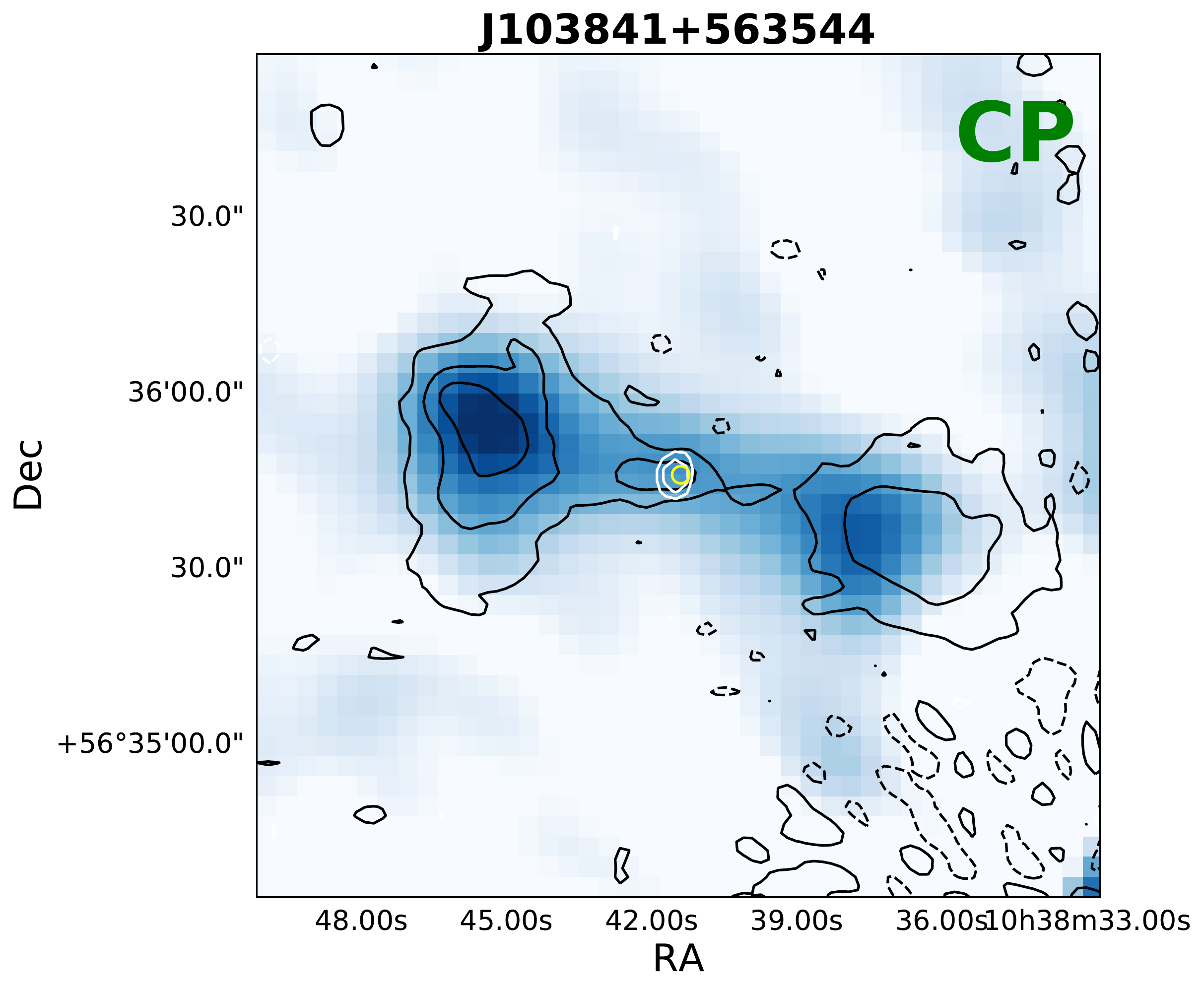}
    \endminipage \hfill    
    \minipage{\textwidth}
        \includegraphics[width=0.24\linewidth] 
            {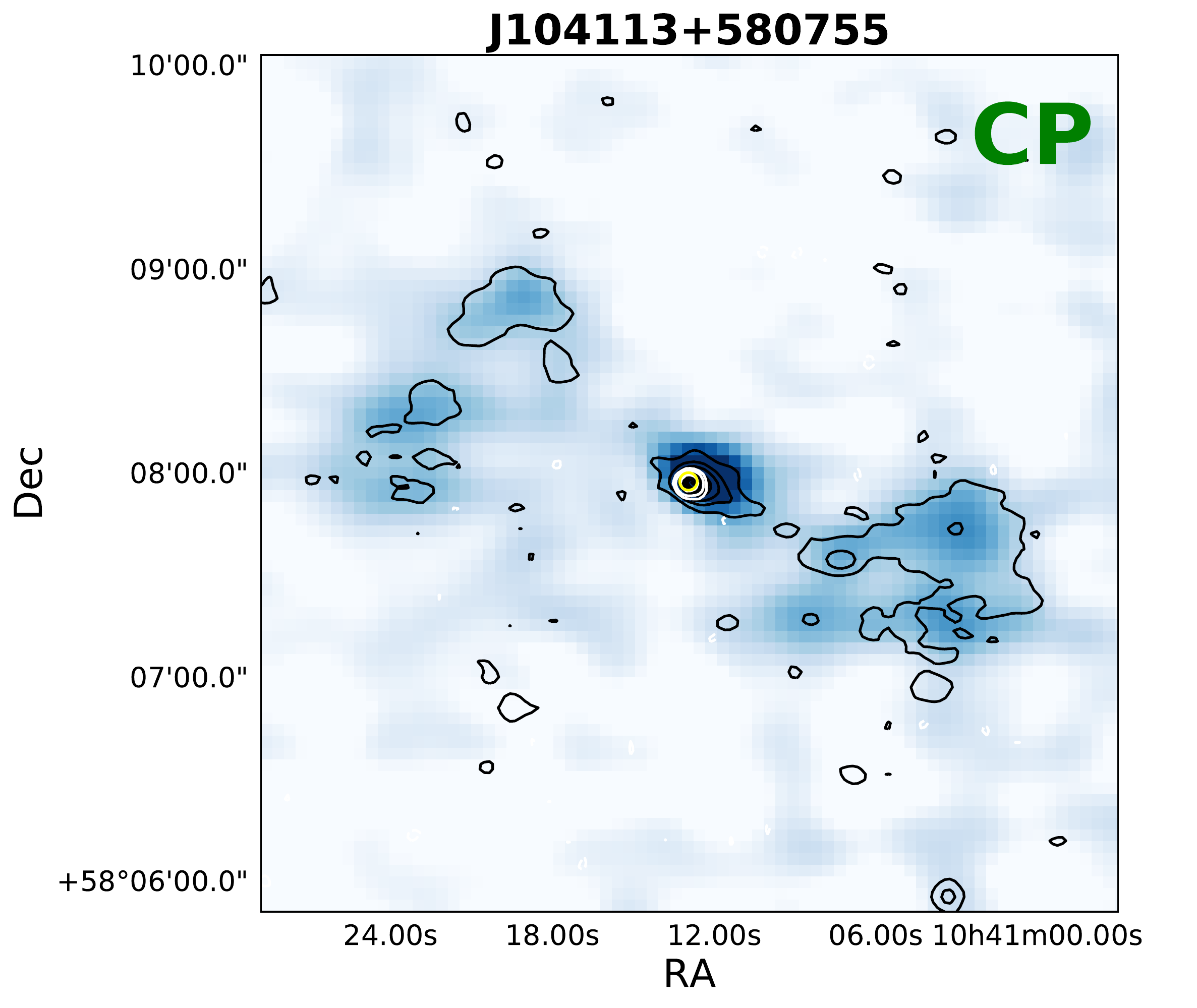}
        \includegraphics[width=0.24\linewidth]
        {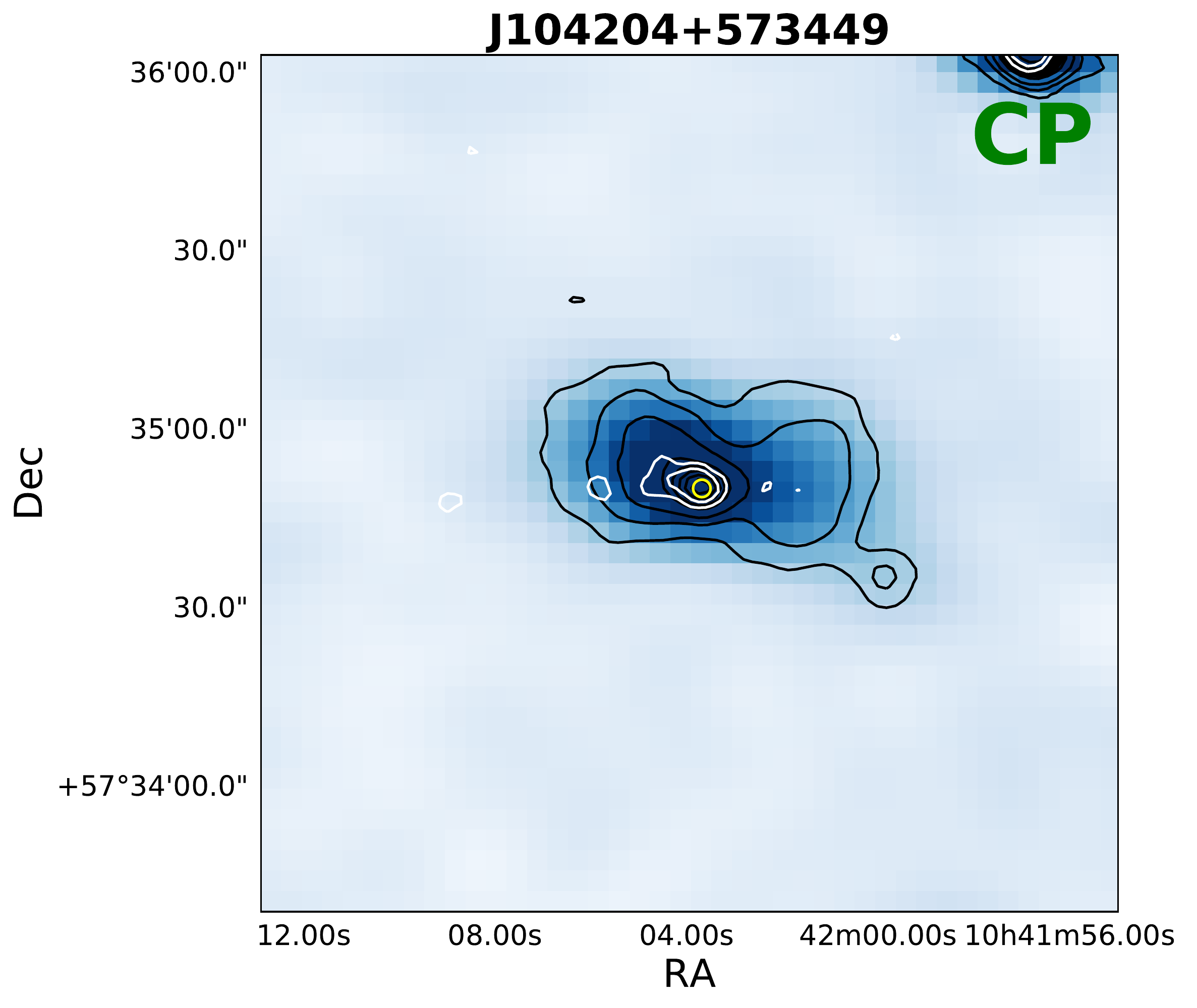} \includegraphics[width=0.24\linewidth]
        {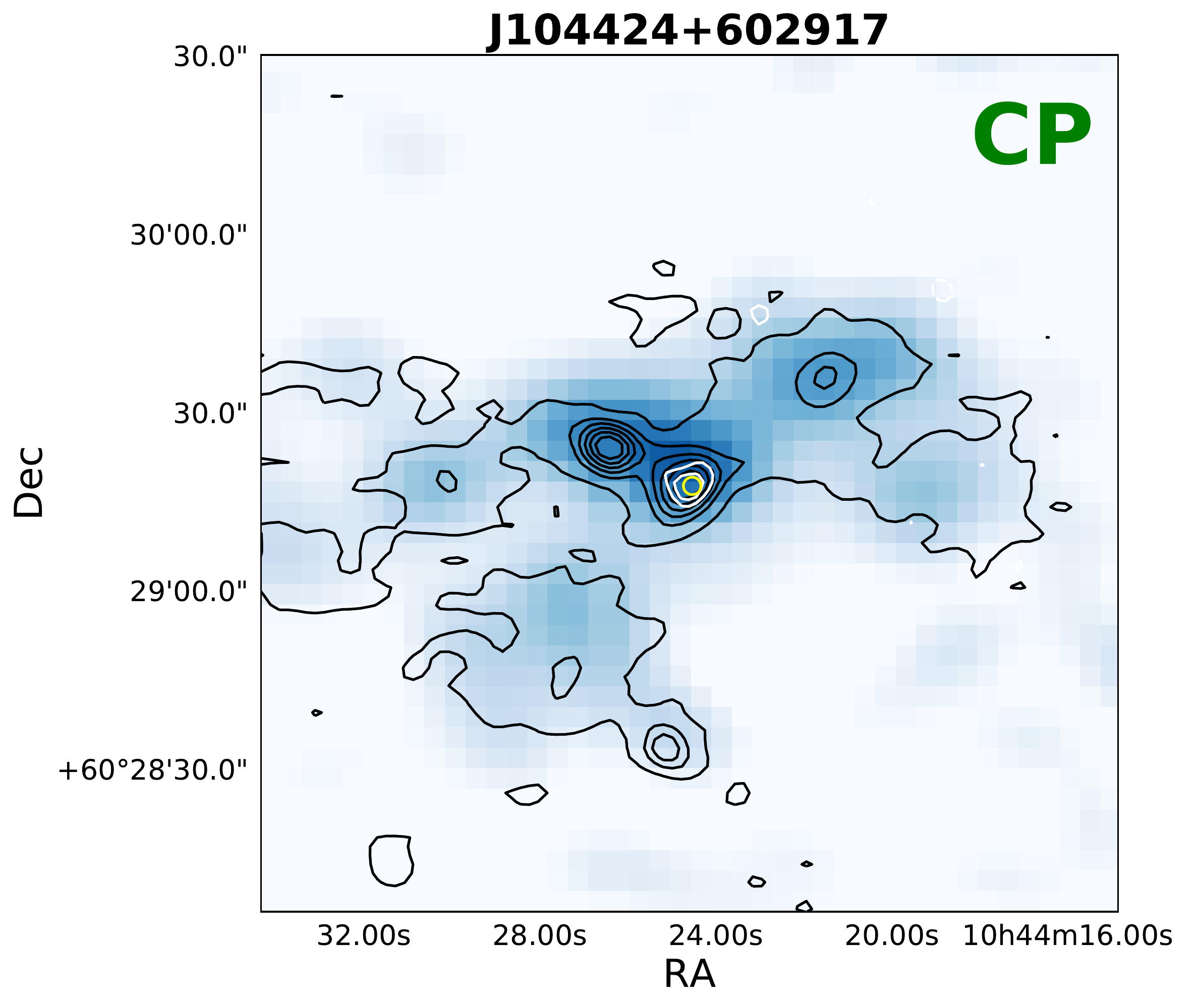}
        \includegraphics[width=0.24\linewidth] 
            {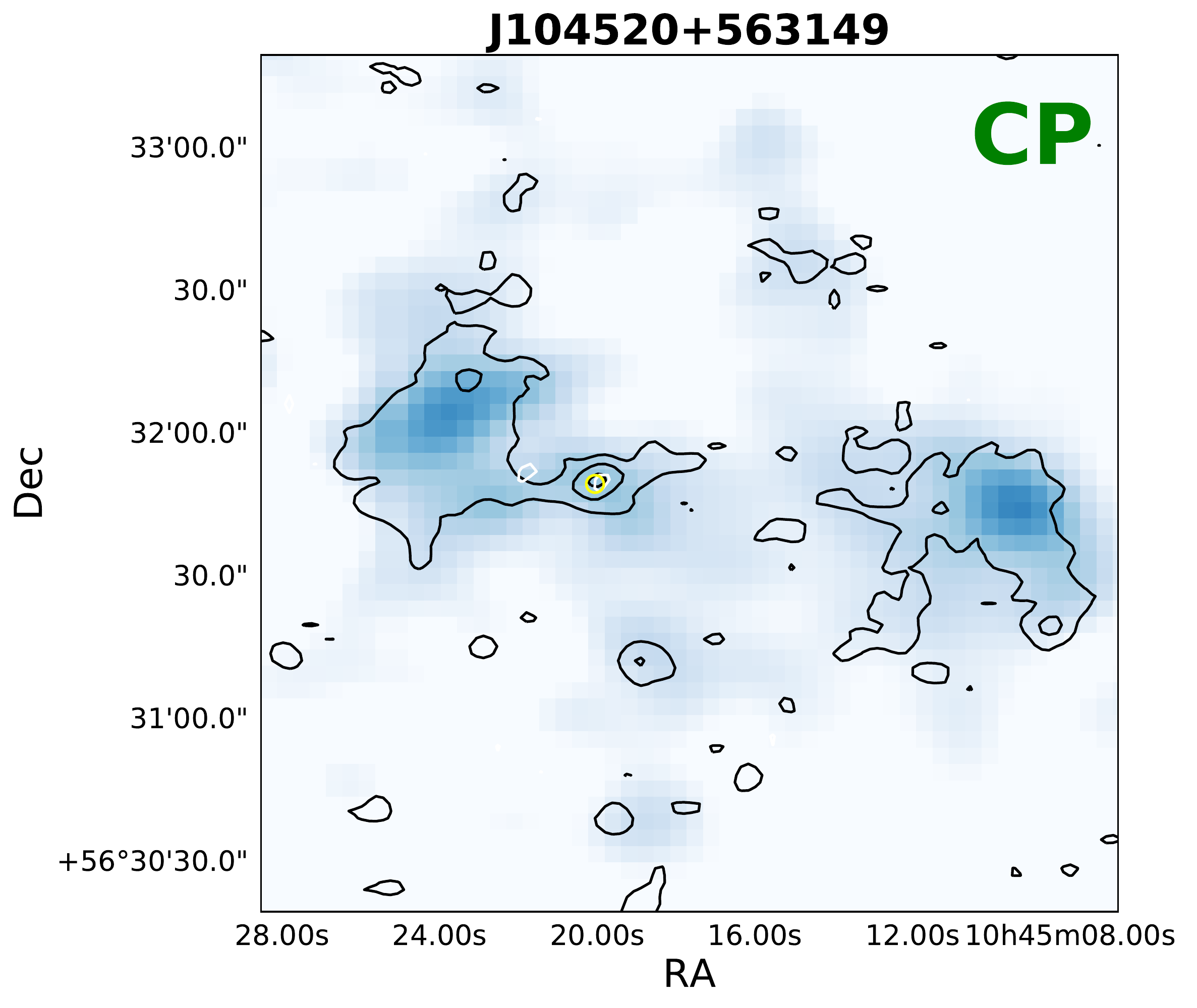}
    \endminipage \hfill
    \minipage{\textwidth}
            \includegraphics[width=0.24\linewidth]
        {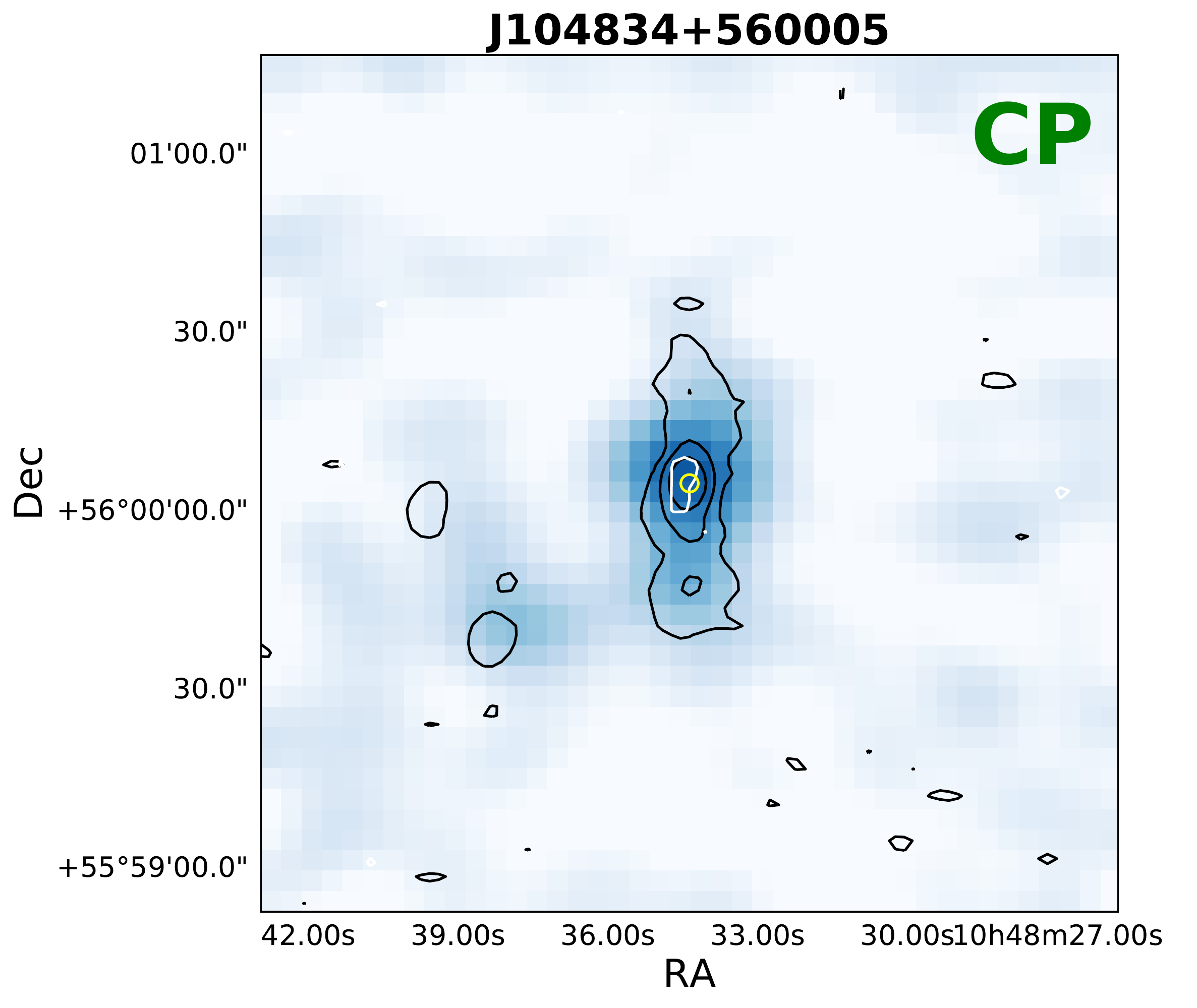}
            \includegraphics[width=0.24\linewidth]
        {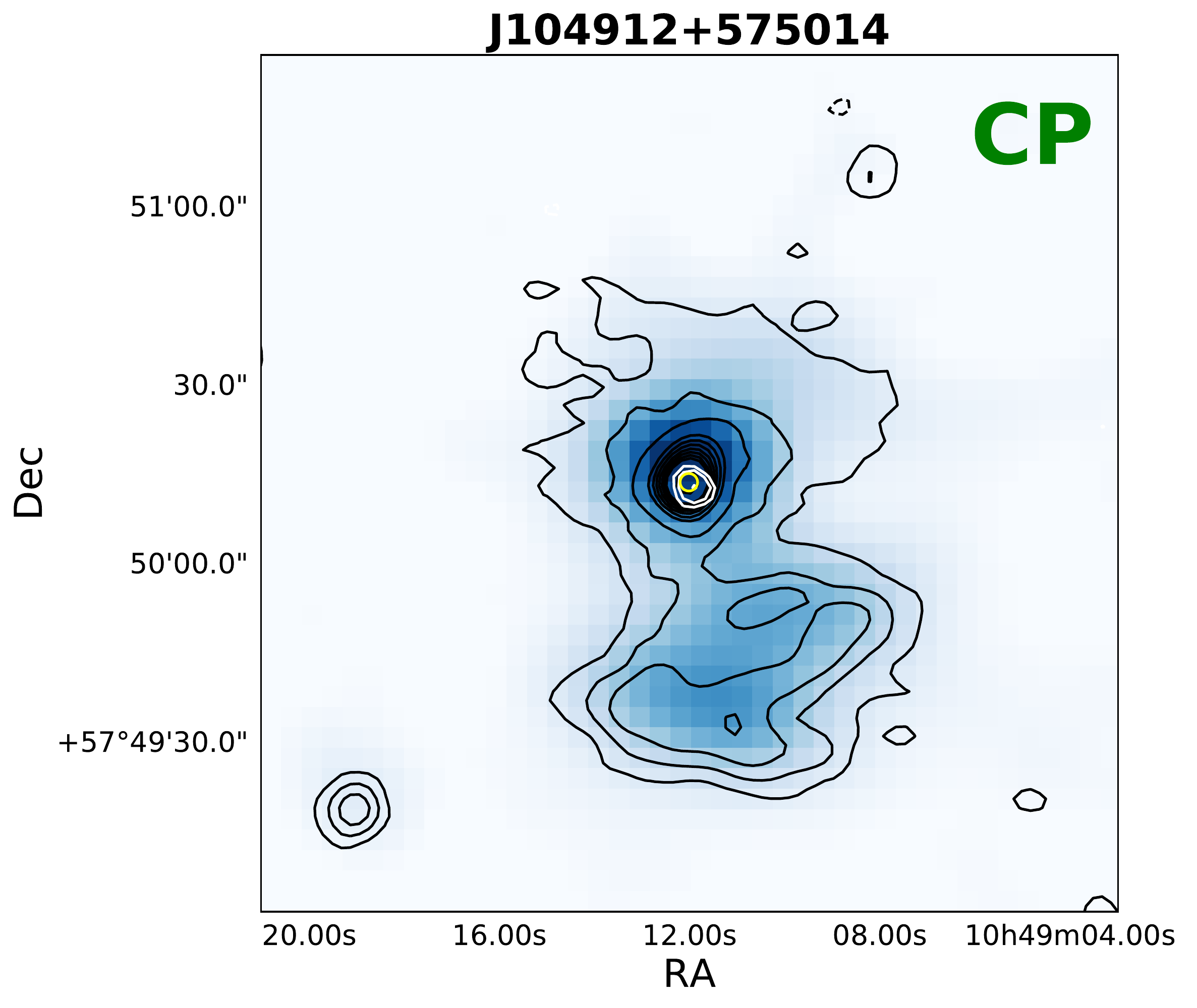}
        \includegraphics[width=0.24\linewidth]
        {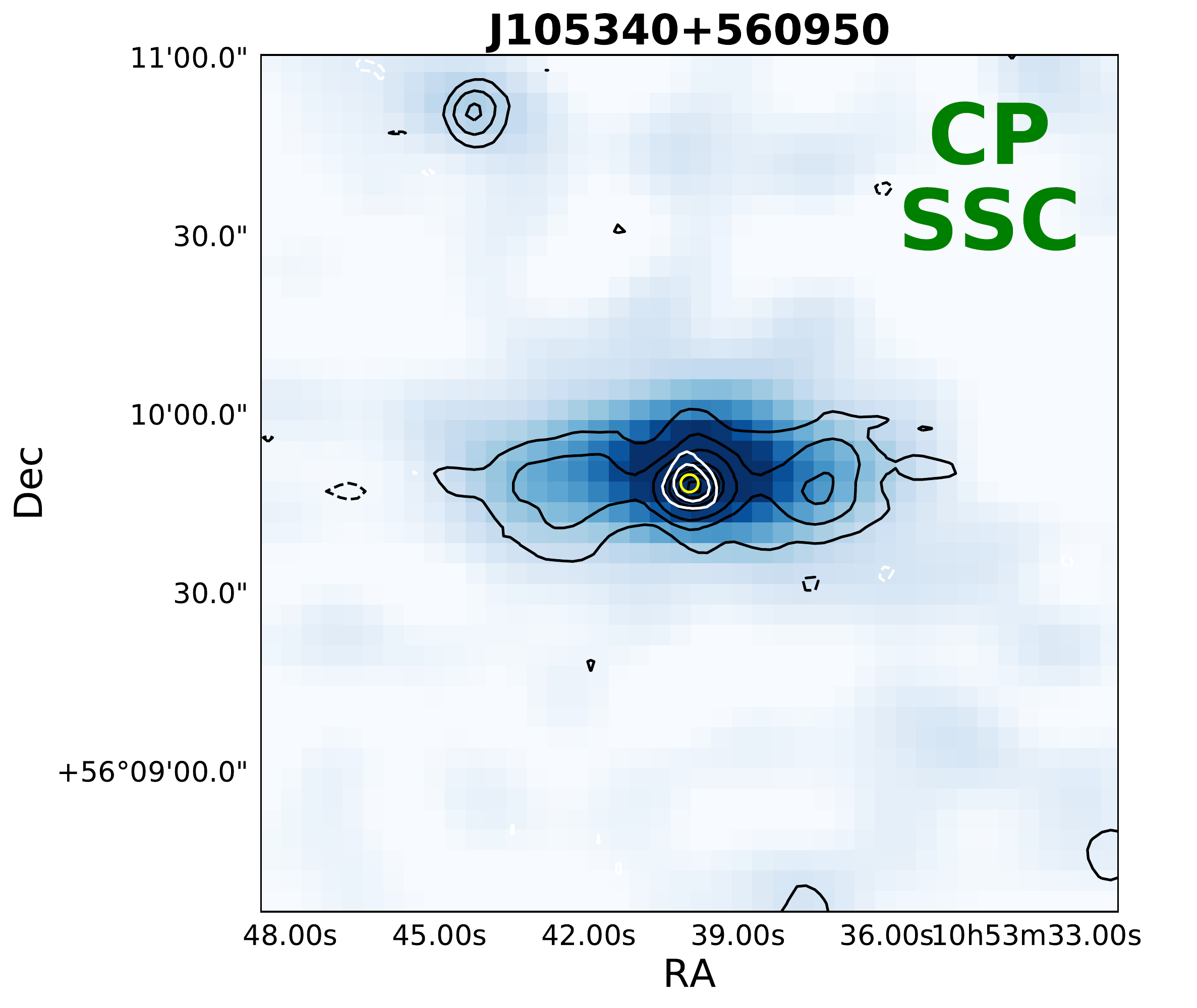} 
        \includegraphics[width=0.24\linewidth]
        {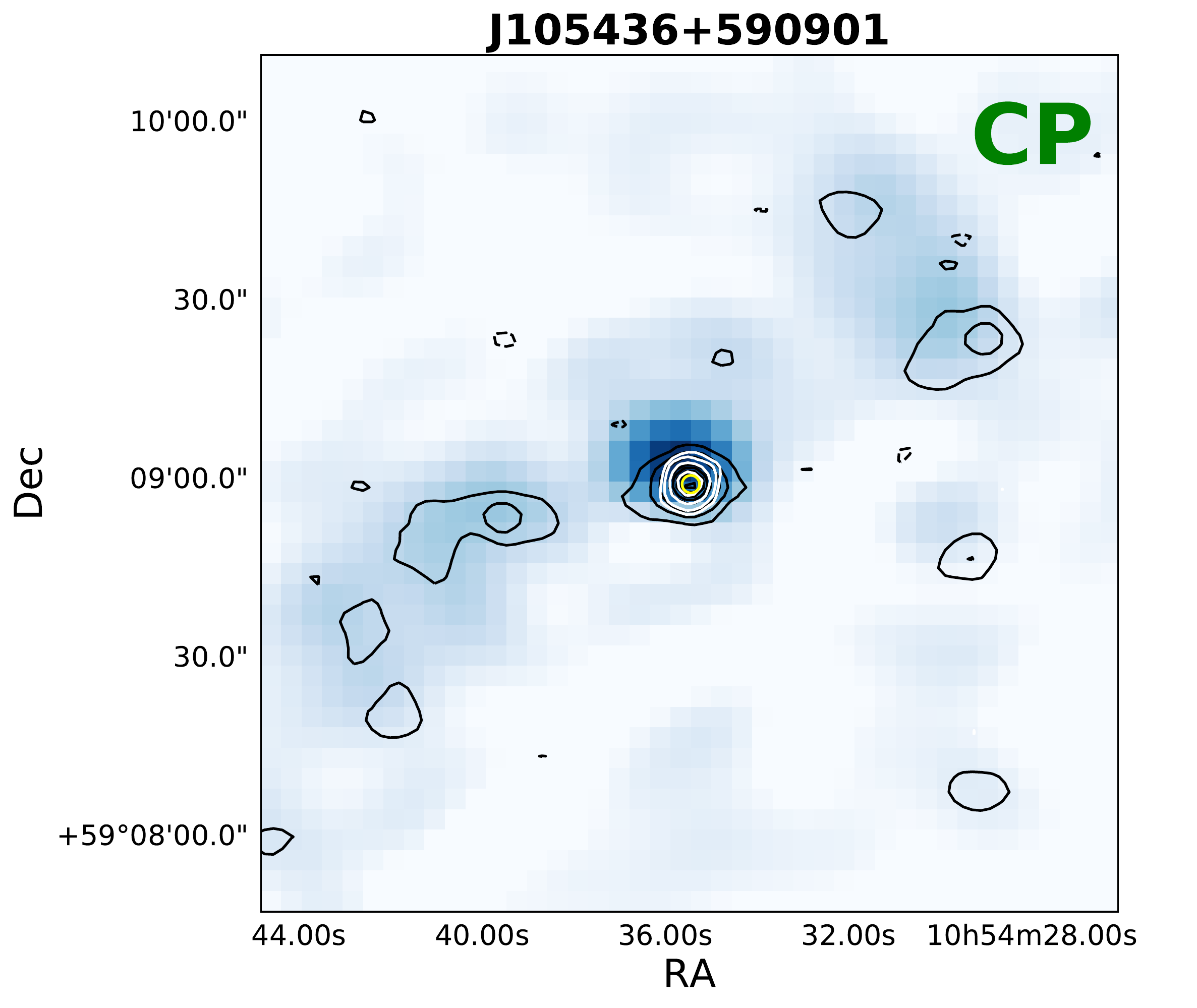}
    \endminipage \hfill
    \minipage{\textwidth}
    \includegraphics[width=0.24\linewidth]
            {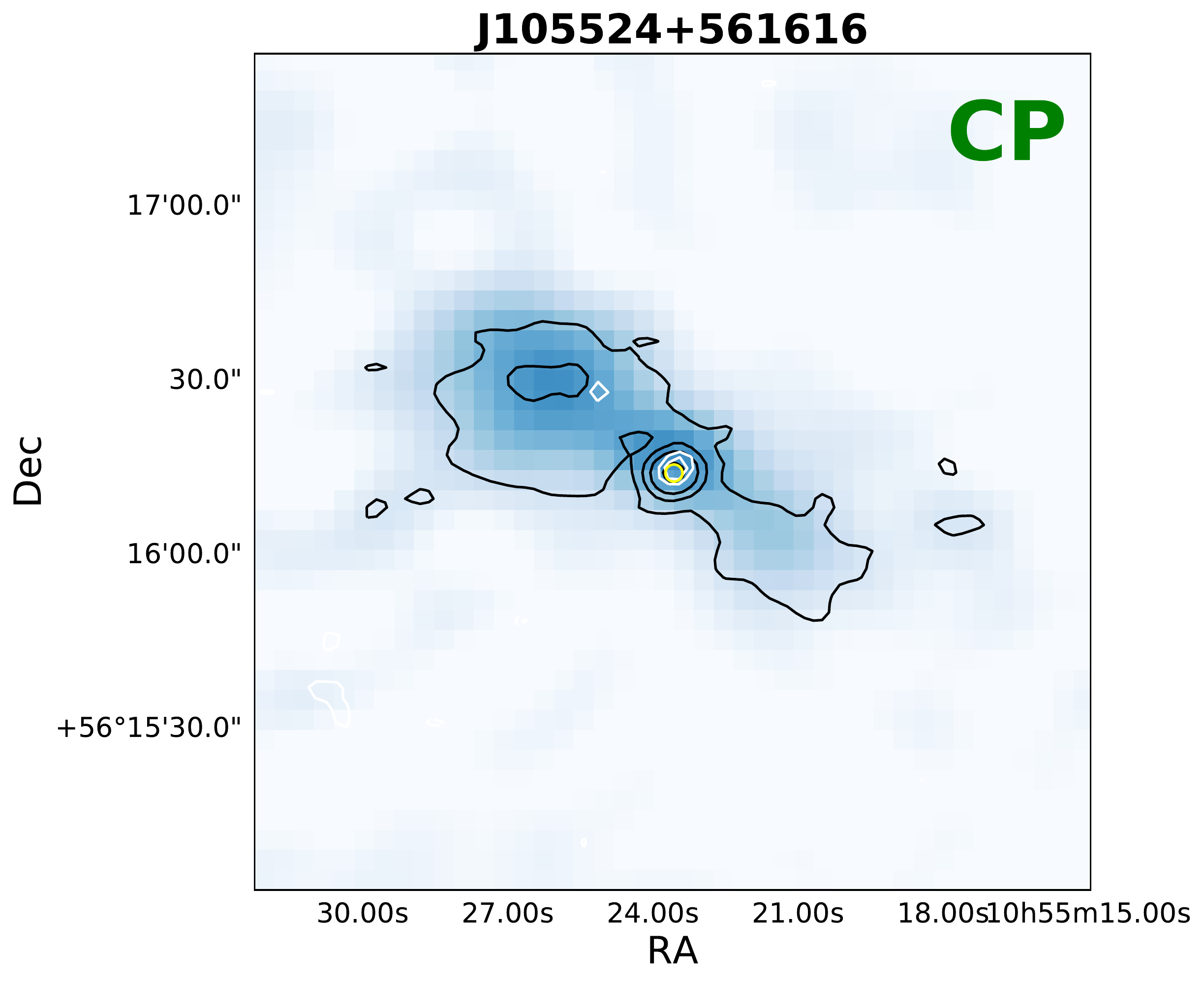}
            \includegraphics[width=0.24\linewidth]
        {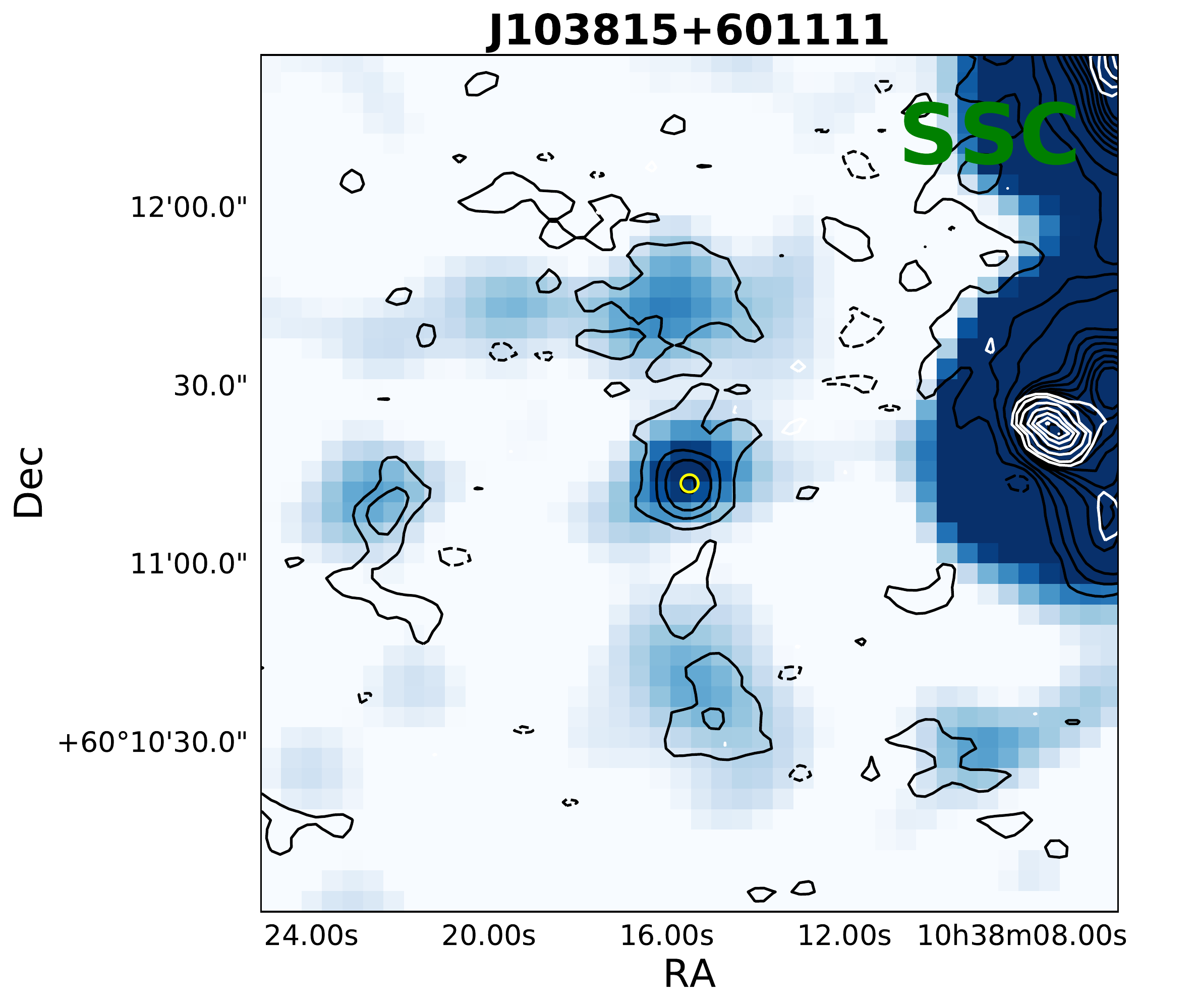}
        \includegraphics[width=0.24\linewidth]
        {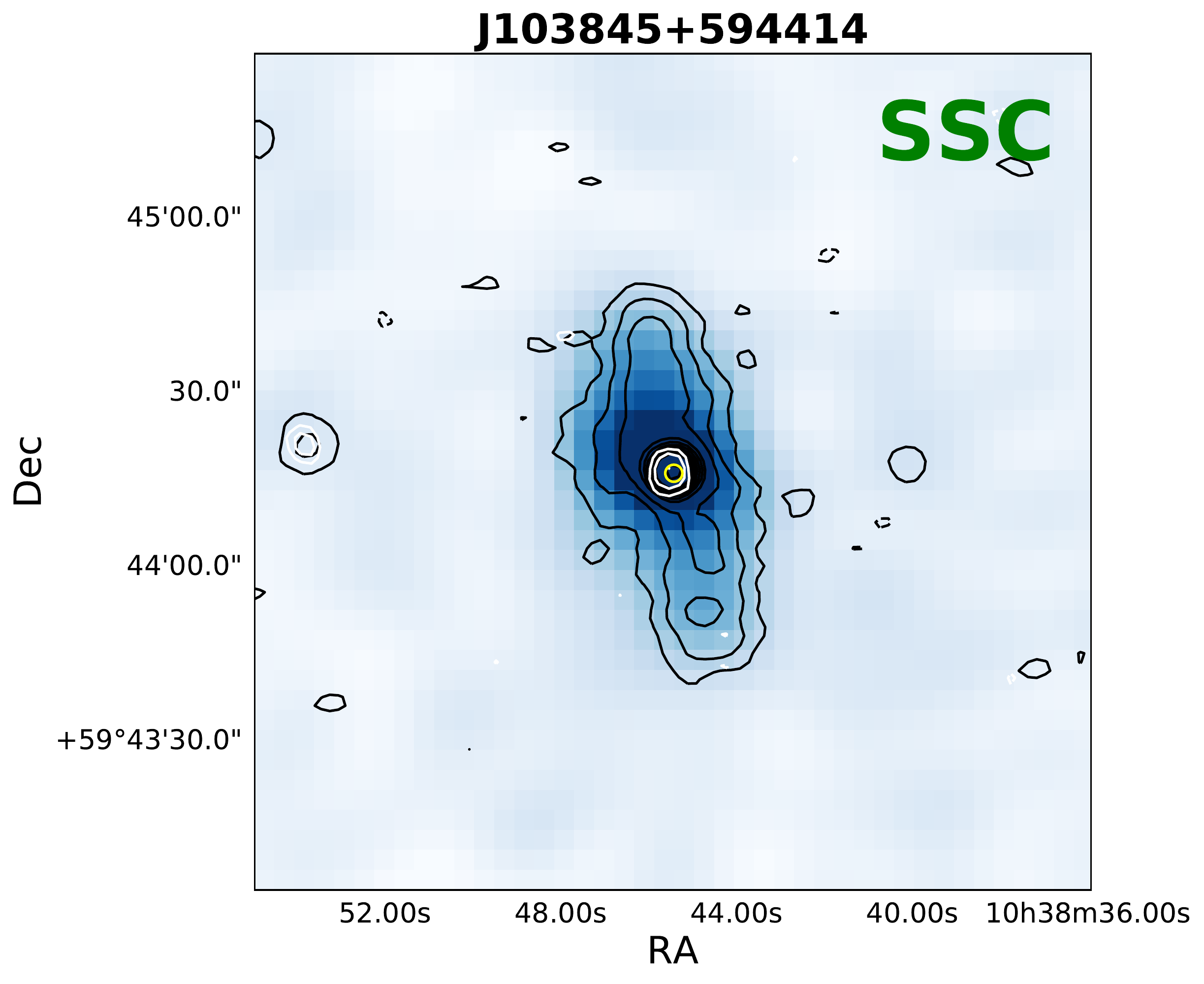} 
        \includegraphics[width=0.24\linewidth]
        {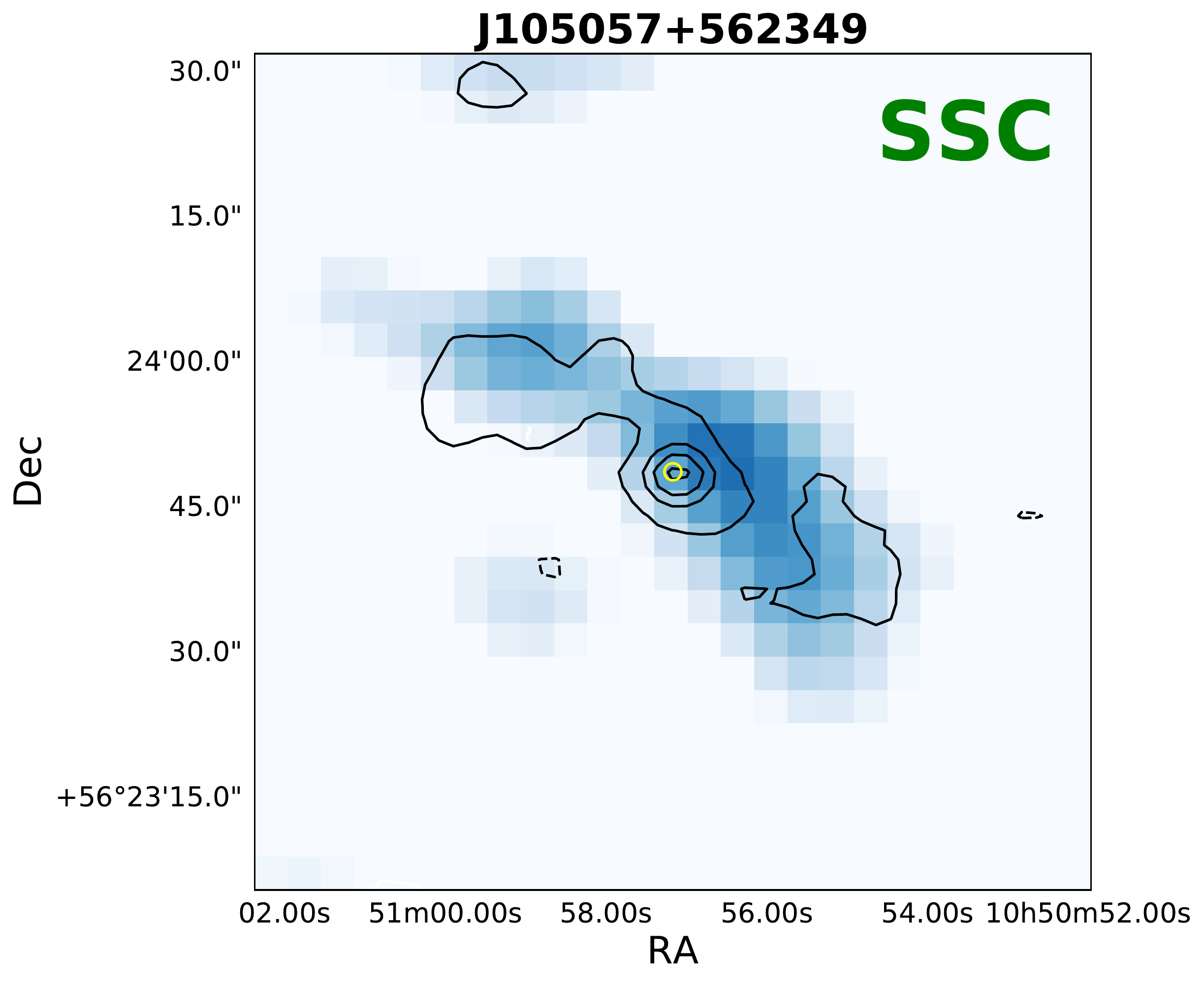}
    \endminipage \hfill
    \minipage{\textwidth}
        \includegraphics[width=0.24\linewidth]
        {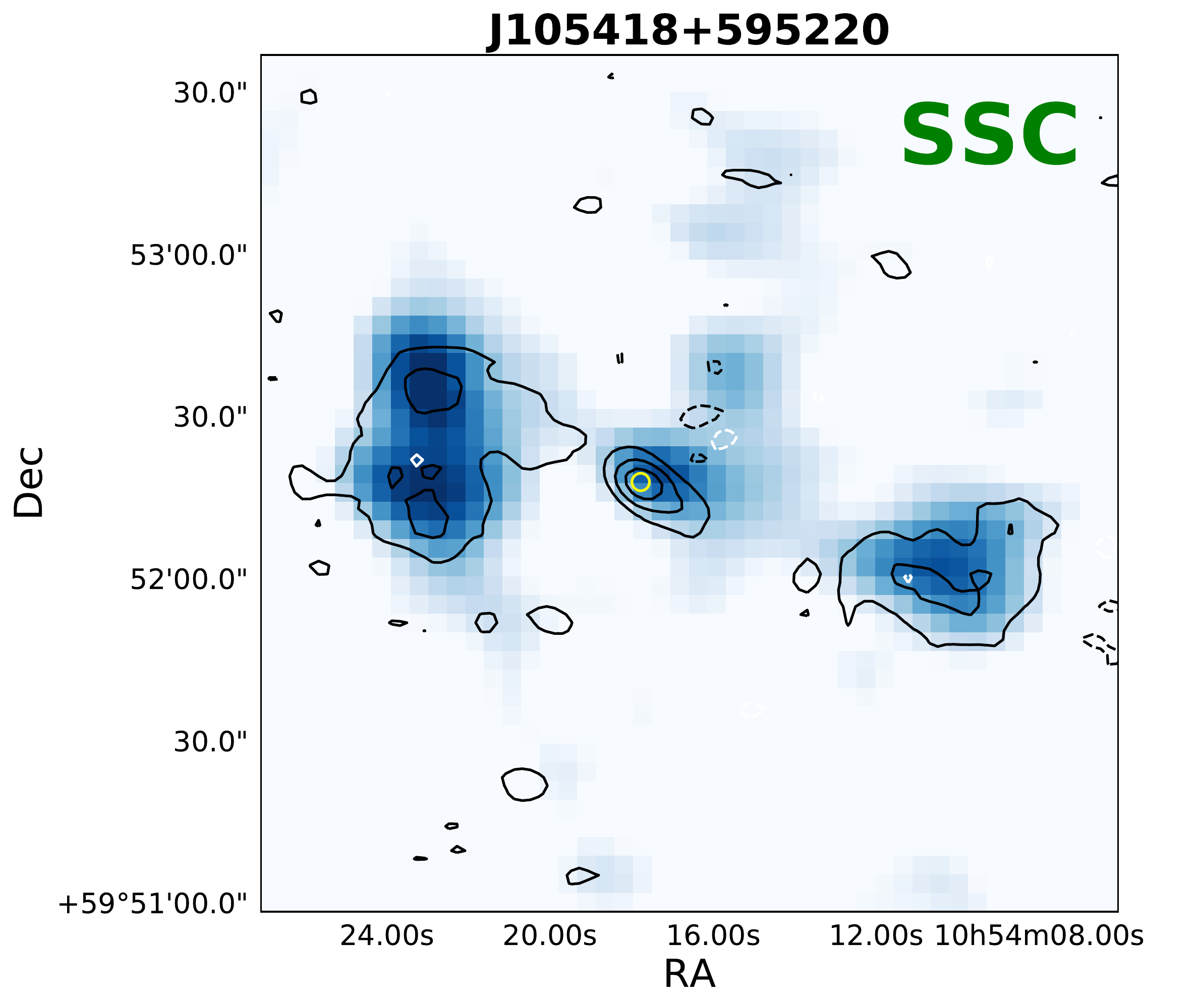}
        \includegraphics[width=0.24\linewidth]
        {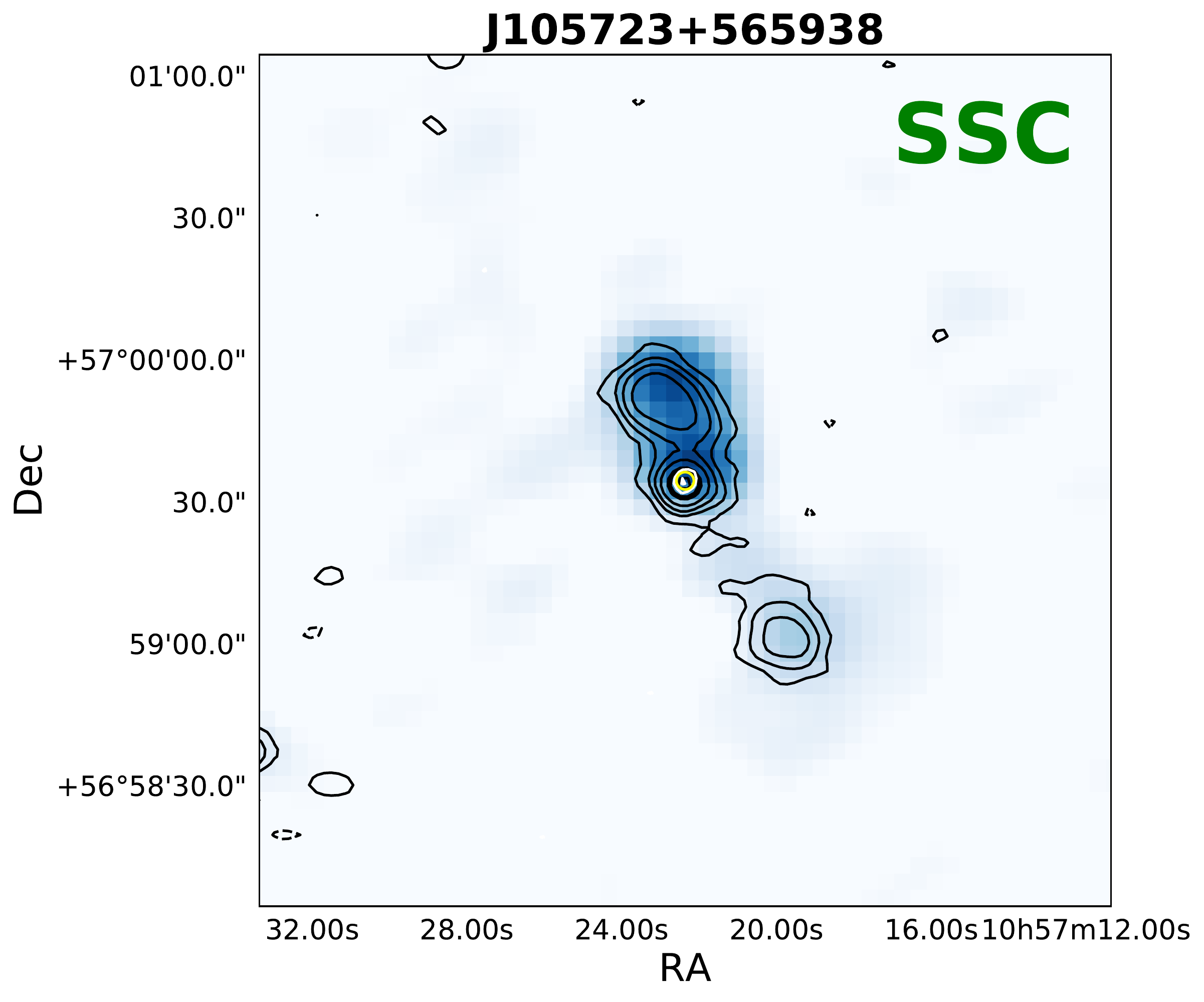}
        \includegraphics[width=0.24\linewidth]
        {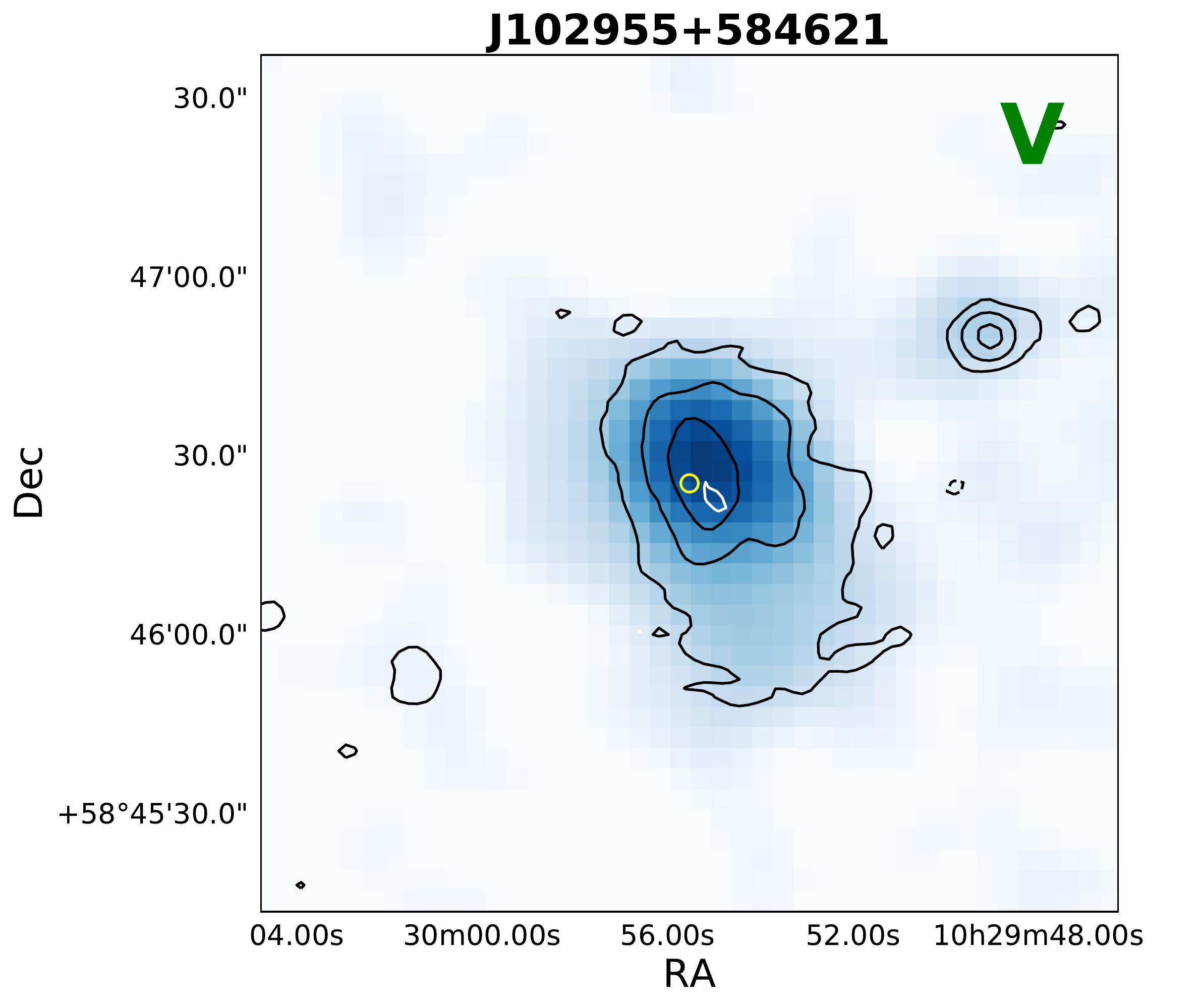}
        \includegraphics[width=0.24\linewidth]
        {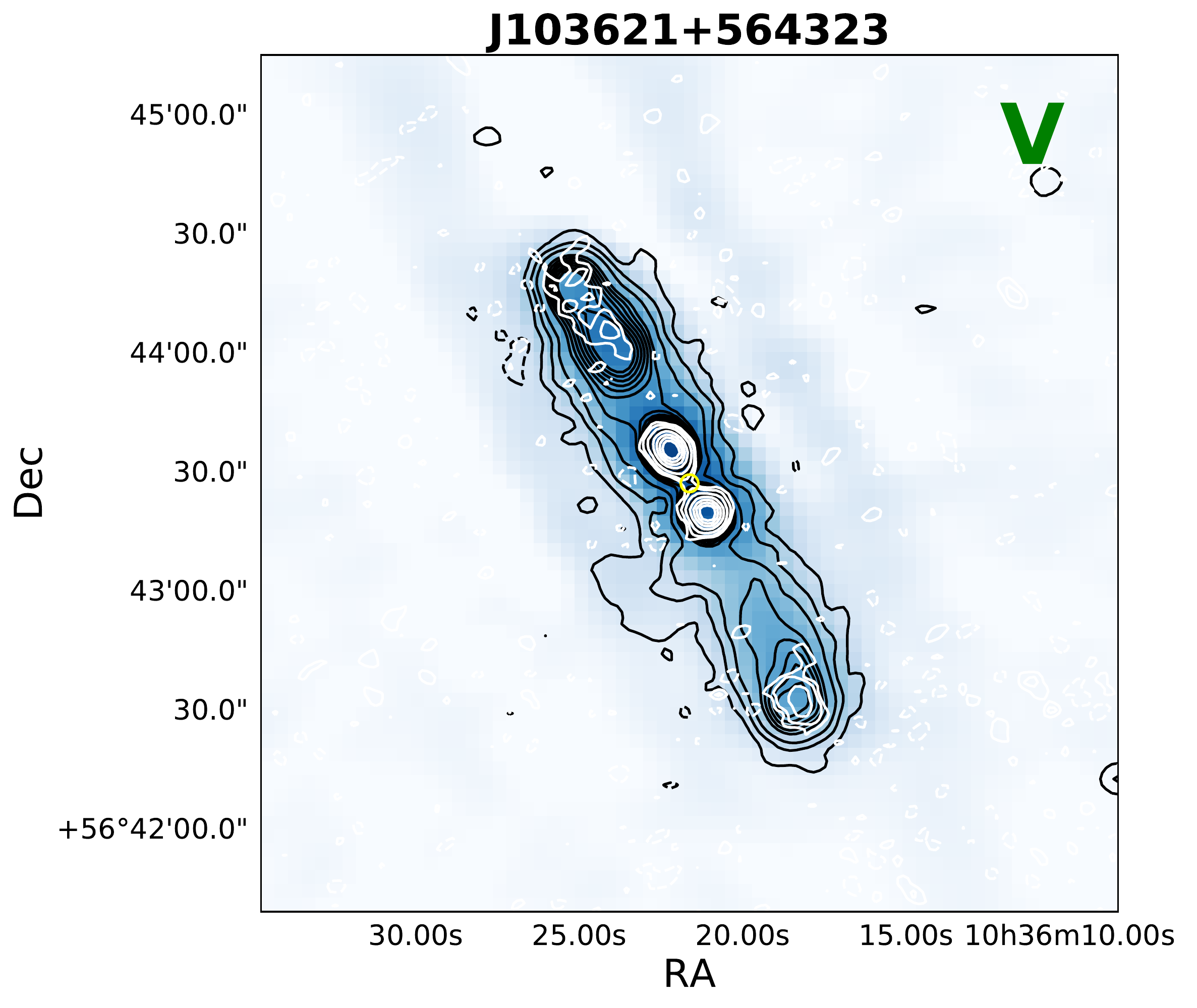}
    \endminipage \hfill
    \minipage{\textwidth}
        \includegraphics[width=0.24\linewidth]
        {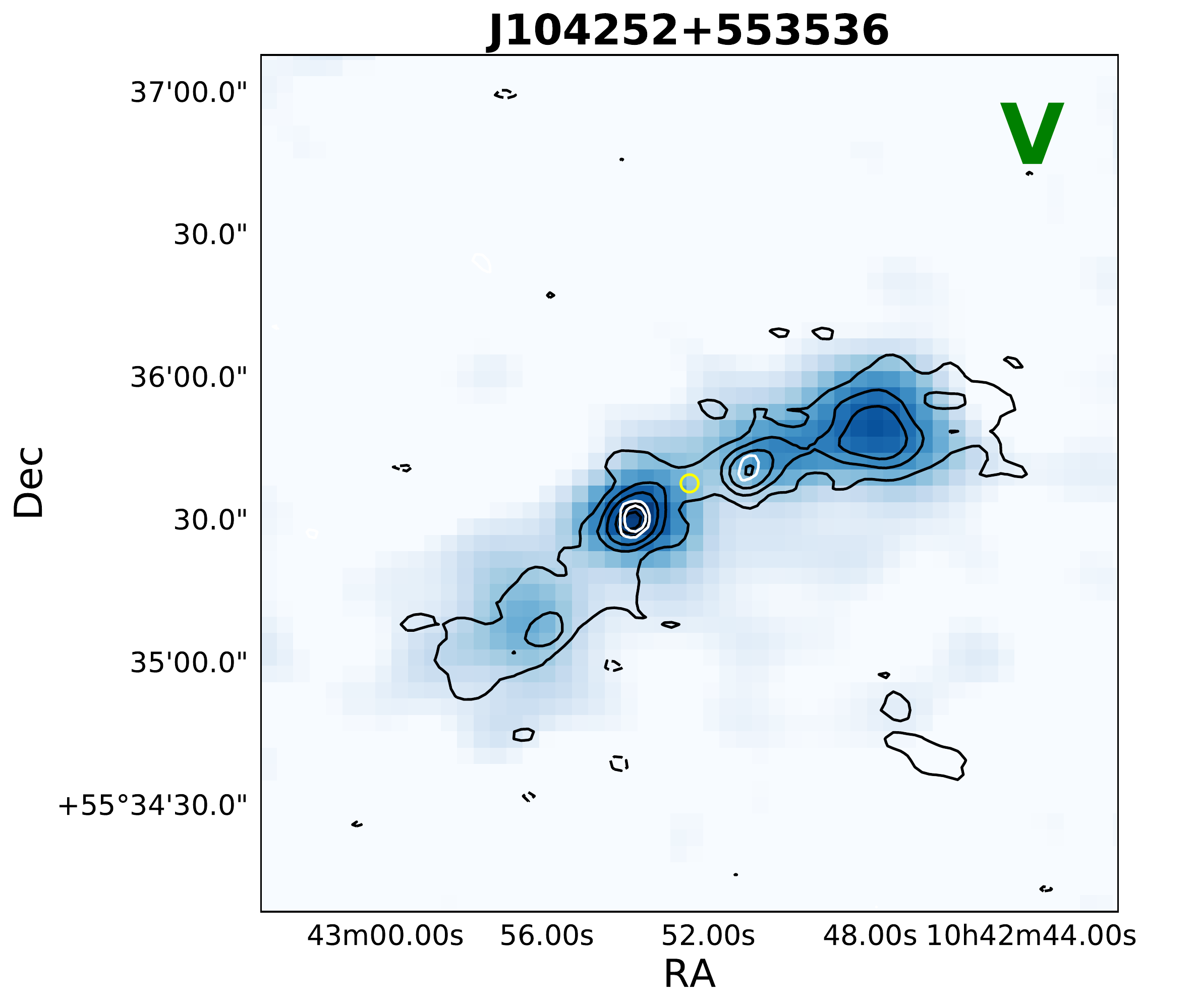}
        \includegraphics[width=0.24\linewidth]
        {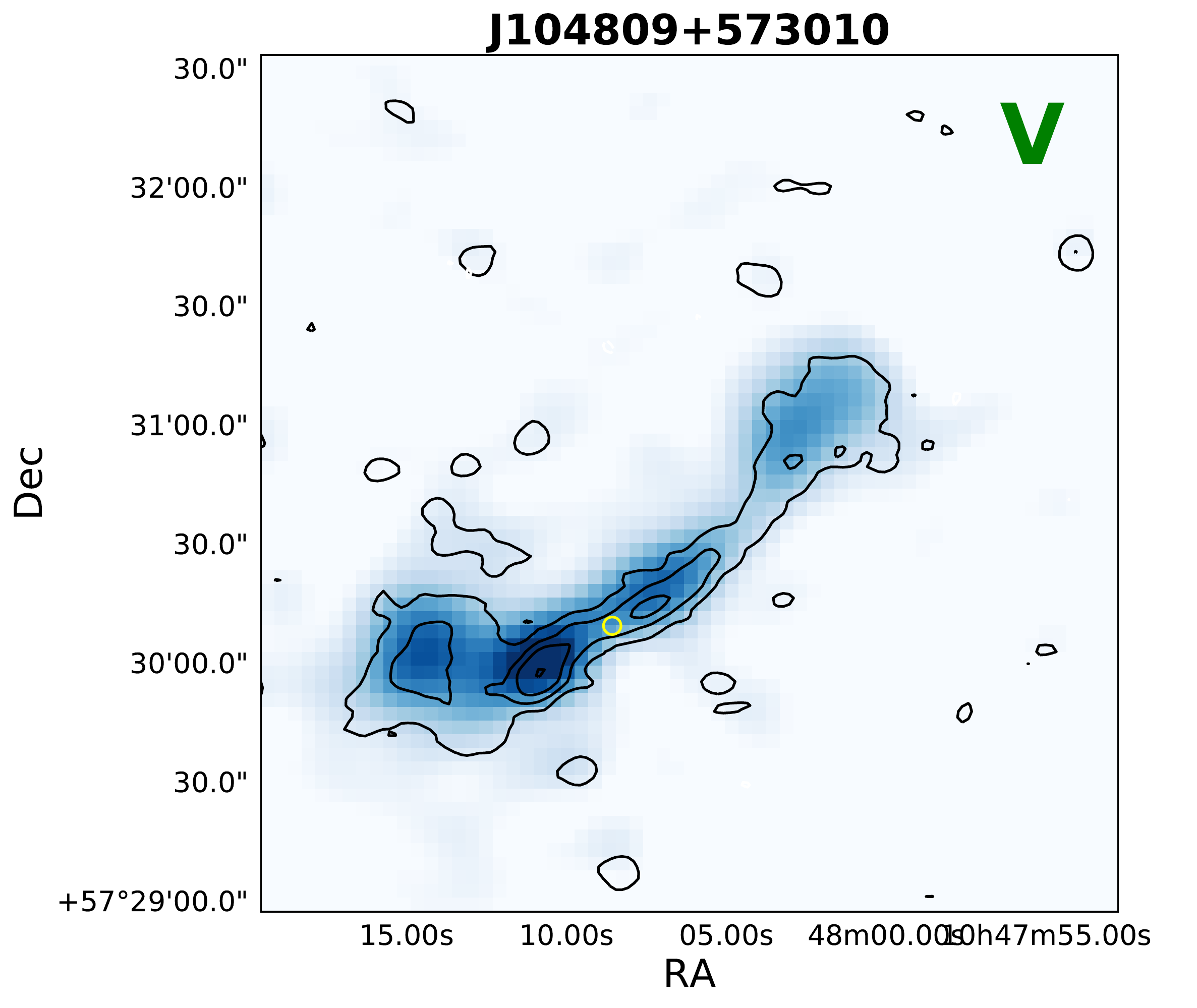}
        \includegraphics[width=0.24\linewidth]
        {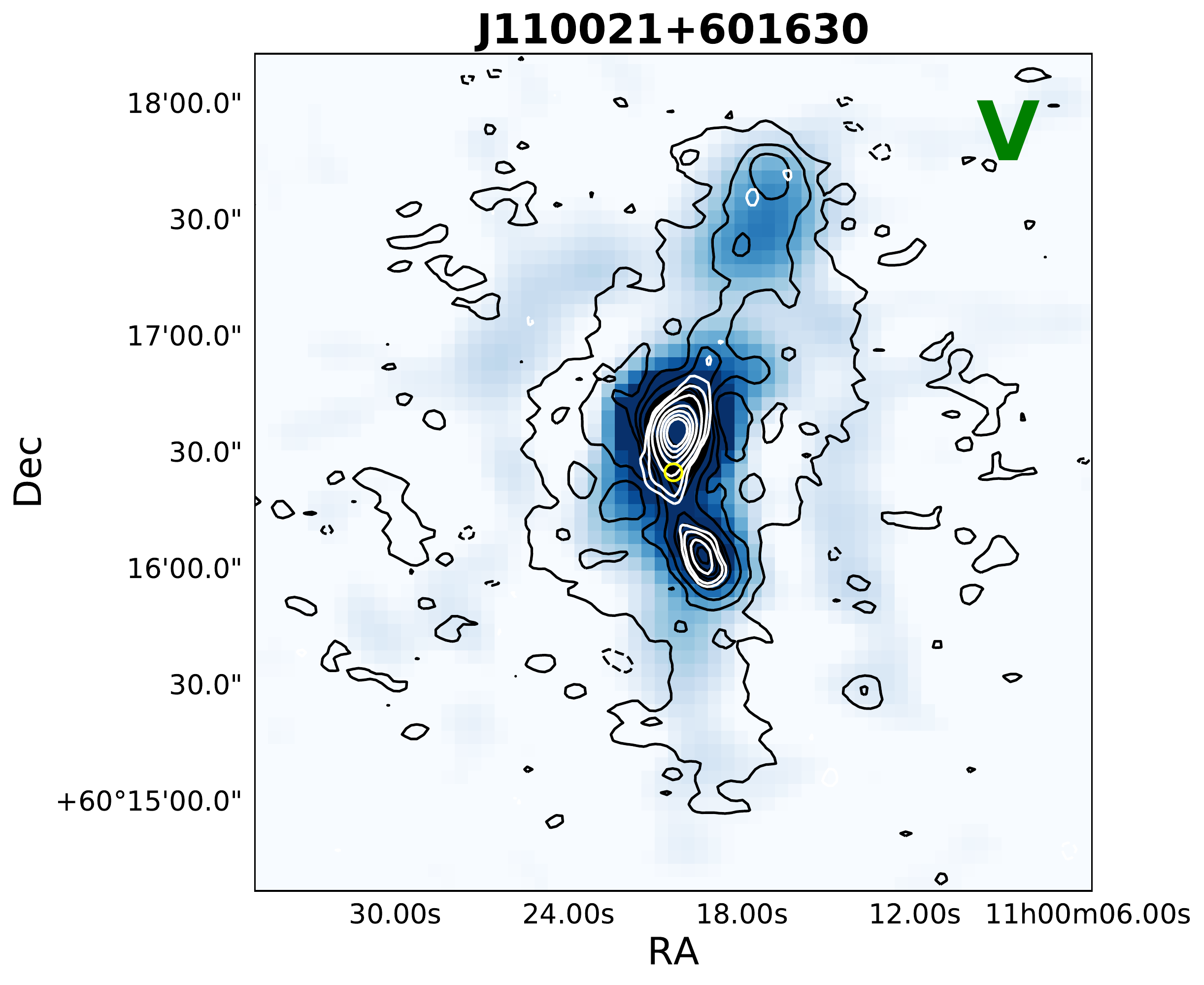}
    \endminipage \hfill
  \caption{Candidate restarted radio galaxies selected on the basis of CP$_{\rm 1400~MHz}$ and low-SB extended emission (see Sect.~\ref{subsec:results_CP}), steep spectrum of the inner region (see Sect.~\ref{subsec:results_SI}), and visual inspection (see Sect.~\ref{subsec:results_visual}). LOFAR radio contours (-3, 3, 5, 10, 20, 30, 40, 50, 100, 150, 200 $\rm \times$ $\rm \sigma_{local}$) of the 6$^{\prime\prime}$ resolution map contours (black) and FIRST contours (white) are overlaid on the LOFAR 18$^{\prime\prime}$ resolution image. The position of the optical counterpart is noted with a yellow circle.}
\label{fig:all_criteria1}
\end{figure*}

\section{Measured and derived parameters of candidate restarted radio galaxies}
Here we present all the derived and measured properties of candidate restarted radio galaxies with their respective uncertainties.

\begin{table*} [h]
\caption{List of candidate restarted radio galaxies. The columns show: source name of the candidate restarted radio source; the core flux density measured directly in the FIRST image; the core flux density measured directly in the LOFAR 6${^{\prime\prime}}$ image; total flux density value from the NVSS catalogue or image (see Sect.~\ref{subsec:results_CP}); radio luminosity of the source, stellar mass and the linear size of the radio galaxy. The asterisk denotes the upper limit values.}
\label{tab:candidate_restarted_appendix}
\centering
\begin{tabular} {l c l l c c c p{5cm}}
\hline\hline
Name & $\rm S_{core, 1400MHz}$ & $\rm S_{core, 150MHz}$ & $\rm S_{tot, 1400MHz}$     & $\rm L_{150MHz}$              & $M_{\star}$ & size \\
& [mJy] & [mJy] & [mJy] & [$\times$ 10$^{25}$ WHz$\rm^{-1}$] & [$\times$ $10^{11}$ $M_{\sun}$]  & [kpc] \\ \hline
J102955+584621 & 0.42*  &   4.33*   &   4.8      &        1.5 $\pm$ 0.4         &    -              &  394 $\pm$ 31 \\
J103416+590523 & 0.61   &   3.43    &   3.3      &        1.3 $\pm$ 0.3         &    13 $\pm$ 13    & 490 $\pm$ 40 \\
J103508+583940 & 0.95   &   4.86    &   1.21     &        1.5 $\pm$ 0.2         &    4 $\pm$ 1  & 401.80 $\pm$ 0.06\\
J103621+564323 & 0.70*  &   62.11*  &   58.3     &        33 $\pm$ 6            &    10 $\pm$ 4     & 1030 $\pm$ 40 \\
J103730+600011 & 0.87   &   3.40    &   2.9      &        0.0022 $\pm$ 0.0003   &    0.14 $\pm$ 0.02& 35.40 $\pm$ 0.02\\
J103815+601111 & 0.40*  &   3.17    &   2.25*    &        0.18 $\pm$ 0.02       &    1.0 $\pm$ 0.2  & 423.89 $\pm$ 0.08\\ 
J103841+563544 & 1.26   &   1.49*   &   1.8      &        2.7 $\pm$ 0.7         &    6 $\pm$ 7      & 670 $\pm$ 40\\
J103845+594414 & 1.52   &   12.97   &   3.1      &        1.2 $\pm$ 0.2         &    1.9 $\pm$ 0.8  & 383 $\pm$ 18\\ 
J104113+580755 & 2.42   &   5.63    &   4.3      &        1.3 $\pm$ 0.1         &    2.1 $\pm$ 0.5  & 1022.8 $\pm$ 0.1\\ 
J104204+573449 & 1.28   &   6.29    &   4.9      &        1.6 $\pm$ 0.2         &    4 $\pm$ 2      &  514.07 $\pm$ 0.06\\
J104252+553536 & 0.28*  &   7.35*   &   7.4      &        2.6 $\pm$ 0.3         &    12 $\pm$ 4     & 824.2 $\pm$ 0.2\\
J104424+602917 & 1.08   &   4.16    &   6.3      &        0.7 $\pm$ 0.2         &    1.0 $\pm$ 0.3  & 378 $\pm$ 29\\
J104520+563149 & 0.59   &   1.27    &   2.4      &        0.002 $\pm$ 0.014     &    0.05 $\pm$ 0.30  & 110 $\pm$ 340\\ 
J104809+573010 & 0.34*  &   1.18    &   2.25*    &        0.80 $\pm$ 0.09       &    2.1 $\pm$ 0.5  & 828.0 $\pm$ 0.1\\ 
J104834+560005 & 0.72   &   3.25    &   2.5      &        1.8 $\pm$ 0.3         &    26 $\pm$ 13    & 450 $\pm$ 8 \\ 
J104912+575014 & 1.57   &   9.66    &   2.6      &        0.15 $\pm$ 0.02       &    2.9 $\pm$ 0.5  & 147.24 $\pm$ 0.04\\ 
J105057+562349 & 0.22*  &   1.51    &   1.16*    &        0.5 $\pm$ 0.1     &    8 $\pm$ 8      &  441 $\pm$ 22\\ 
J105340+560950 & 1.27   &   7.73    &   3.0      &        >1.13    &    > 11.9         & > 697 \\
J105418+595220 & 0.29*  &   1.94    &   2.25*    &        2.7 $\pm$ 0.4    &    3 $\pm$ 1      & 972 $\pm$ 17\\ 
J105436+590901 & 4.38   &   4.64    &   5.2      &        2.5 $\pm$ 0.3    &    -              &  1046.9 $\pm$ 0.1 \\ 
J105524+561616 & 1.14   &   4.42    &   3.9      &        0.5 $\pm$ 0.1   &    3 $\pm$ 1      & 419 $\pm$ 26   \\ 
J105723+565938 & 0.77   &   10.93   &   1.38*    &        >6.26    &    >16.6          &  >769 \\ 
J110021+601630 & 3.37*  &   41.14*  &   60.8     &        2.5 $\pm$ 0.3    &    1.1 $\pm$ 0.2  & 608.03 $\pm$ 0.07    \\ \hline \hline 
\end{tabular}   
\end{table*}        

\end{document}